\documentclass[10pt,journal,compsoc]{IEEEtran}
\ifCLASSINFOpdf
\else
\fi
\usepackage{graphicx}
\usepackage[table,xcdraw]{xcolor}
\usepackage{multirow}
\usepackage{amsmath,amssymb,amsfonts}
\usepackage{cite}
\usepackage{enumerate}
\usepackage[shortlabels]{enumitem}
\usepackage[detect-all]{siunitx}
\usepackage{tabularx}
\usepackage{colortbl}
\usepackage{comment}
\usepackage[justification=centering]{caption,subcaption}
\usepackage{algpseudocode}
\usepackage{algorithmicx}
\usepackage{graphicx}
\usepackage[ruled, linesnumbered]{algorithm2e}
\usepackage{listings}
\usepackage{tikz}
\usepackage{ragged2e}
\usepackage[table]{xcolor}
\usepackage{multirow}
\newcommand*\circled[1]{\tikz[baseline=(char.base)]{
\node[shape=circle,draw,inner sep=0.7pt] (char) {#1};}}
\usepackage{makecell}

\usepackage{footnote}
\makesavenoteenv{tabular}
\makesavenoteenv{table}
\makeatletter
\algnewcommand{\LineComment}[1]{\State \(\triangleright\) #1}
\usepackage{amssymb}% 
\usepackage{pifont}% 
\usepackage[normalem]{ulem}

\definecolor{britishracinggreen}{rgb}{0.0, 0.26, 0.15}
\newcommand{\cmark}{\color{britishracinggreen}\ding{51}}%
\newcommand{\xmark}{\color{red}\ding{55}}
\newcommand{\hlt}[1]{{\color{black} #1}}
\newcommand{\hltr}[1]{{\color{black} #1}}

\newcommand{\SubItem}[1]{
    {\setlength\itemindent{15pt} \item[-] #1}
}
\begin{document}
\title{Snoopy: A Webpage Fingerprinting Framework with Finite Query Model for Mass-Surveillance}

\author{Gargi Mitra,
        Prasanna Karthik Vairam,
        Sandip Saha,
        Nitin Chandrachoodan,
        V. Kamakoti% <-this % stops a space
\IEEEcompsocitemizethanks{\IEEEcompsocthanksitem Gargi Mitra, Sandip Saha, Nitin Chandrachoodan and V. Kamakoti are from Indian Institute of Technology Madras, India.\protect\\ E-mail: \{gargim@cse, CS20S044@cse, nitin@ee, kama@cse\}.iitm.ac.in\protect\\
Prasanna Karthik Vairam is from National University of Singapore. E-mail: prasanna@comp.nus.edu.sg
}
% \thanks{Manuscript received April 19, 2005; revised August 26, 2015.}
}

\IEEEtitleabstractindextext{%
\begin{abstract}
\justifying
Internet users are vulnerable to privacy attacks despite the use of encryption. Webpage fingerprinting, an attack that analyzes encrypted traffic, can identify the webpages visited by a user. Recent research works have been successful in demonstrating webpage fingerprinting attacks on individual users, but have been unsuccessful in extending their attack for mass-surveillance. The key challenges in performing mass-scale webpage fingerprinting arises from (i) the sheer number of combinations of user behavior and preferences to account for, and; (ii) the bound on the number of website queries imposed by the defense mechanisms (e.g., DDoS defense) deployed at the website. These constraints preclude the use of conventional data-intensive ML-based techniques.  

In this work, we propose Snoopy, a first-of-its-kind framework, that performs webpage fingerprinting for a large number of users visiting a website. Snoopy caters to the generalization requirements of mass-surveillance while complying with a bound on the number of website accesses (finite query model) for traffic sample collection. For this, Snoopy uses a feature (i.e., sequence of encrypted resource sizes) that is either unaffected or predictably affected by different browsing contexts (OS, browser, caching, cookie settings). Snoopy uses static analysis techniques to predict the variations caused by factors such as header sizes, MTU, and User Agent String that arise from the diversity in browsing contexts.
We show that Snoopy achieves $\approx90\%$ accuracy when evaluated on most websites, across various browsing contexts. 
% For cases where Snoopy did not perform well, we use a simple ensemble of Snoopy and an ML-based technique that achieves $\approx 97\%$ accuracy while adhering to the finite query model.
A simple ensemble of Snoopy and an ML-based technique achieves $\approx 97\%$ accuracy while adhering to the finite query model, in cases when Snoopy alone does not perform well.
%for improving the prediction accuracy in the few scenarios where the accuracy of Snoopy alone was comparatively low. Such an ensemble is useful for achieving a high prediction accuracy.
% while complying to the constraints of a limited number of queries.
\end{abstract}

}

\maketitle

% is not selected <OR> if conference mode is selected - because compsoc
% conference papers position the abstract like regular (non-compsoc)
\IEEEdisplaynontitleabstractindextext

% \fi
\IEEEpeerreviewmaketitle

\ifCLASSOPTIONcompsoc
\IEEEraisesectionheading{\section{Introduction}\label{sec:intro}}
\else
\section{Introduction}\label{sec:intro}
\fi

Leakage of private information is one of the biggest concerns for Internet users today. Recent reports~\cite{leak1,leak2} suggest that sensitive information that could cause imminent personal harm to Internet users, including banking passwords, salary details, health records, location information, and CCTV footage, have been leaked in the dark web.
% \hlt{\sout{several times over the past few years}}. 
% Examples of such personal information include banking passwords, salary details, health records, location information, and CCTV footage. 
While privacy loss for personal information is easily perceivable, it is not so obvious for other types of information. For example, the Cambridge Analytica scandal showed that the political leanings of an ordinary individual may not be worthy to the attackers, but the political alignment of a larger demography can help them predict the outcome of an election~\cite{cambridge}. Such incidents show that some user information that are seemingly unimportant to an individual might inadvertently turn sensitive when collected on a \textit{mass scale}. 

One of the largest sources of mass-scale information about personal preferences of Internet users are their web browsing activities~\cite{seo2000learning,middleton2001capturing}. Information such as the {\em identity of websites} visited by users from a demography could be useful to an attacker, for instance, to gauge the popularity of websites in the region. However, more fine-grained information such as the {\em identity of webpages} (a \textit{website} can have many \textit{webpages}) visited by the users on a targeted website could be much more useful to the attackers. For instance, a surge in the number of visitors to the ``fixed-deposit" page of a bank website shortly after the announcement of a new policy is critical to the bank. This information, if leaked, could help competing banks estimate its growth and also design counter-marketing strategies. In addition, a malware may be placed on a popular webpage, that could potentially infect every visitor of the webpage. In this paper, we focus on designing a {\em mass-surveillance method} that can potentially reveal such \textit{fine-grained} sensitive information from a large number of Internet users based on their {\em web browsing activities}. 

\begin{figure}[t]
    \centering
    \includegraphics[scale=0.15]{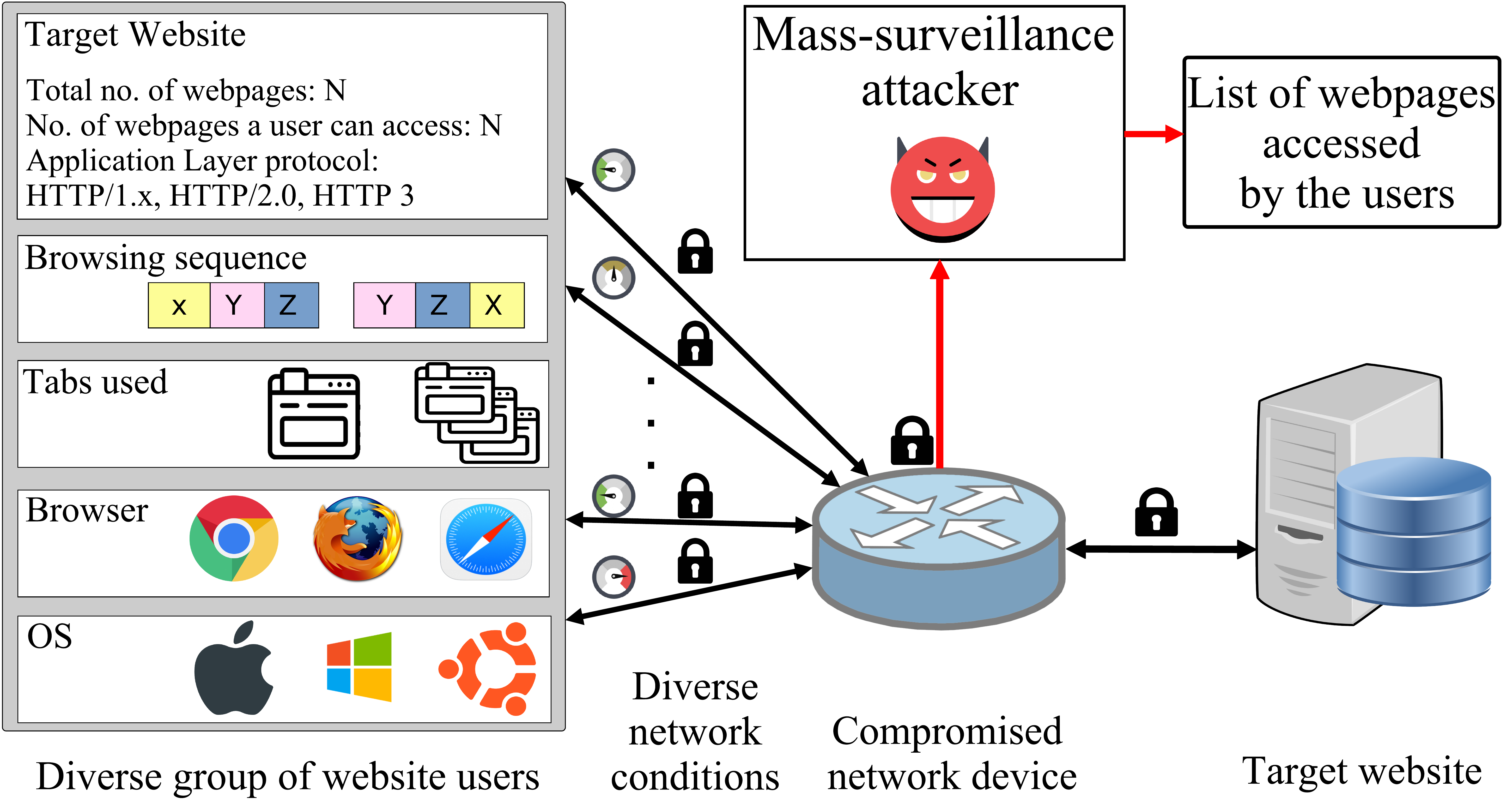}
    % \caption{A High Level Overview of Snoopy\vspace{-15pt}}
    \caption{Snoopy, the proposed framework for webpage identification attack on a mass-scale}
    \label{fig:snoopy-hlov}
    \vspace{-15pt}
\end{figure}

Straightforward methods to identify the webpages visited by the users of a website include compromising the end-user devices and the websites' servers (e.g., extracting decryption key, installing fake certificates). However, these methods are highly intrusive and are easily detectable, making them unsuitable for long-term surveillance. Note that the importance of information grows manifold when collected over a longer period of time~\cite{goel2012does,olejnik2012johnny,muller2014hats,bird2020replication}, since it ensures sufficient coverage of users and also reflects changing trends in their behavior.
Encrypted Traffic Analysis (ETA)~\cite{danezis2010traffic,chapman2011automated,cai2012touching,miller2014know,gu2015novel,hayes2016k,panchenko2016website,xu2018multi,yan2018feature,alan2019client,ghiette2020scaling,shen2020fine,wang2021s} (specifically, webpage fingerprinting) is the most promising non-intrusive method to collect such information for long periods of time by merely capturing the encrypted traffic exchanged between the websites' server and the end-users. The webpages are identified by formulating signatures that can identify them uniquely when accessed by users through encrypted channels.
%ETA is performed at a network device on the encrypted traffic (e.g., HTTPS~\cite{}, VPN~\cite{}, Tor~\cite{} traffic) exchanged between the websites' server and the end-user using webpage fingerprinting techniques. %The attacker employs statistical techniques on the encrypted traffic to reveal either the website or the webpage (within the website) accessed by a user. In this work, we focus on {\em webpage} identification (using webpage fingerprinting) since fine-grained information is more valuable.
%Although ETA has been widely used for inferring coarse-grained information (i.e., for identifying website accessed by a user), fewer research works have attempted to unravel fine-grained information such as the {\em webpage} (within a website) accessed by a user (i.e., webpage fingerprinting attacks). 
%While the existing works on webpage fingerprinting~\cite{danezis2010traffic,chapman2011automated,cai2012touching,miller2014know,gu2015novel,hayes2016k,panchenko2016website,xu2018multi,yan2018feature,alan2019client,ghiette2020scaling,shen2020fine,wang2021s} have been highly successful on a small scale, they are not suitable for handling the complexities of mass-scale analysis even with just HTTPS (i.e., without Tor/VPN).
While several existing webpage fingerprinting attacks~\cite{danezis2010traffic,chapman2011automated,cai2012touching,miller2014know,gu2015novel,hayes2016k,panchenko2016website,xu2018multi,yan2018feature,alan2019client,ghiette2020scaling,shen2020fine,wang2021s} target particular users, our goal is to target a large number of users visiting a particular website, i.e., mass surveillance. Figure~\ref{fig:snoopy-hlov} shows a high-level objective of this paper.

Existing techniques~\cite{danezis2010traffic,chapman2011automated,cai2012touching,miller2014know,gu2015novel,hayes2016k,panchenko2016website,xu2018multi,yan2018feature,alan2019client,ghiette2020scaling,shen2020fine,wang2021s} for webpage fingerprinting are not suitable for mass-scale traffic analysis even with just HTTPS (i.e., without Tor/VPN). 
%Practical mass-surveillance requires a {\em generalized model} of the user traffic to account for a large number of combinations/possibilities of user behavior and network parameters. 
Practical mass-surveillance requires a {\em generalized model} that can identify webpages irrespective of the \textit{browsing context} of a user. Note that, a browsing context refers to the user behavior (such as browser/OS used, number of parallel browser tabs used, etc.) and network parameters during a browsing session.
Building a generalized model that accounts for all possible browsing contexts is challenging. The primary hindrance in the generalization of existing works is the massive number of data points required, i.e., the number of traffic samples required per webpage to account for all possible browsing contexts. For instance, building a generalized model for $3$ operating systems and $3$ browsers will require $9$ times the traffic samples as compared to one-OS-one-browser scenario~\cite{shen2020fine}. Moreover, for achieving the accuracy promised by DL/ML algorithms, collecting adequate number of traffic samples per scenario is also crucial~\cite{sirinam2018deep,shen2020fine}. 
Additionally, any change in the website also requires fresh collection of traffic samples. Therefore, generalizing existing techniques requires assuming a query model that allows the adversary innumerable accesses to the website within a short span of time. However, in practice, we observed that the DDoS/DoS mitigation schemes at the websites detect and block IP addresses that make such attempts. Furthermore, IP addresses that cause anomalies in website analytics, including those involved in attempts to sample webpages repeatedly, are identified and blocked.
% \begin{figure}[t]
%     \centering
%     \includegraphics[scale=0.14]{IEEEtran/Snoopy_finiteqm.pdf}
%     % \caption{A High Level Overview of Snoopy\vspace{-15pt}}
%     \caption{Infinite vs. finite query models}
%     \label{fig:finitequery}
%     \vspace{-15pt}
% \end{figure}
Therefore, to work around it, prior works have assumed knowledge about one or more of the following, thereby compromising generalization: 
\begin{enumerate}
    \item \textbf{User interests} -- Existing works~\cite{danezis2010traffic,chapman2011automated,cai2012touching,miller2014know,ghiette2020scaling} assume that the user is only interested in a subset of webpages. In mass-surveillance, however, the adversary needs to account for a wide variety of user interests;
    \item \textbf{User behavior} -- Most of the existing works~\cite{cai2012touching,gu2015novel,xu2018multi} assume restricted user behavior such as single-tab browsing~\cite{cai2012touching,shen2020fine,wang2021s} and traversing only a limited number of webpages per session~\cite{ghiette2020scaling}. A few works account for the use of more than one tab~\cite{gu2015novel,xu2018multi}, but they were not successful beyond $2$ browser tabs. However, studies~\cite{dubroy2010study} have shown that such assumptions are not valid in practical scenarios. Accounting for multi-tab browsing requires webpage-sequence based data collection, which in turn increases the traffic samples exponentially;
    \item \textbf{User preference} -- Existing works~\cite{miller2014know,hayes2016k,wang2021s} assume prior knowledge about the OS and the browser used, as well as knowledge about browser caching (on/off) and cookie values. However, mass-surveillance necessitates accounting for various combinations of OS and browser configurations; and, 
    % Recent works~\cite{shen2020fine} that aimed to generalize for multiple OS/Browser configurations attempted to do so at the cost of collecting vast amounts of training data from different OS/Browser combinations. While collecting such huge amount of traffic samples is possible for search engines and social media websites (as shown in these works), we found that most business websites prohibit such data collection due to security concerns ; and,
    \item \textbf{Network conditions} -- Existing works build prediction models for a particular network condition, which cannot be generalized to other network conditions. However, mass surveillance requires the prediction model to work across geographical locations and service providers.
\end{enumerate}
% \hlt{\sout{Therefore, for effective mass-scale surveillance, there is a need for a webpage fingerprinting technique that complies to a {\em finite query model}, yet meets the generalization requirements of mass surveillance.} }
% {\em To the best of our knowledge, ours is the first work to attempt webpage fingerprinting attack for mass-scale surveillance}. 
 
Na\"ive attempts at generalization without the addition of adequate number of data points for covering additional scenarios result in poor prediction accuracy. For instance, we witnessed a drastic drop in prediction accuracy (e.g., from 96\% to 76\% in case of Wfin~\cite{yan2018feature}) of existing ML-based works when we broadened the user interest from a small number of webpages\footnote{We conducted these experiments on a popular bank website.} to a large number (refer to Figure~\ref{fig:userinterest} in Section~\ref{sec:user-interests}). Further, we also noticed a drop in prediction accuracy (e.g., from 78\% to 58\% in case of OPS~\cite{cai2012touching,chapman2011automated}) when the number of training samples is reduced from $10$ to $3$ per webpage (refer to Figure~\ref{fig:nq-vs-fa} in Section~\ref{sec:fqm}). Likewise, when we tried to generalize existing works in terms of user preference (caching on or off), we found a drop in accuracy (as shown in Table~\ref{tab:browser-config} in Section~\ref{sec:browser-config}) when a model built for one scenario (caching on) was used to predict webpages accessed from a different scenario (caching off). Therefore, availability of a limited number of training samples imposes restrictions on the use of existing DL/ML based techniques.

In our work, we propose \textit{Snoopy}, a practical webpage fingerprinting framework, that meets the generalization requirements of mass-surveillance, while assuming a \textit{finite query model}. With limited number of traffic samples, accounting for the numerous scenarios encountered in mass-surveillance is challenging. Snoopy uses domain knowledge about the transport and application protocols to collect traffic samples in a focused manner. For instance, Snoopy uses {\em encrypted web resource size}, a simple feature used in ETA, that remains largely unaffected by changes in network conditions, eliminating the need to cover this scenario altogether. For generalizing across other scenarios Snoopy relies on static analysis of web object sizes, HTML code and headers to estimate the expected fingerprints. 
In this paper, we analyze the capability of a simple model such as ours in terms of generalization in cases where there are practical constraints for using ML/DL techniques. \hlt{We show that Snoopy was able to achieve more than $90\%$ accuracy for most of the websites we considered, when tested on traffic samples from a diverse set of browsing contexts.} For the few websites where Snoopy achieved comparatively lower accuracy ($\approx 80\%$), we show that it is possible to improve the accuracy to as high as $97\%$ by using an ensemble of Snoopy and an ML-based technique, that complies with the constraints of a finite query model. 
% Add result summary. Further, we also explore how our method can complement existing ML/DL techniques to make them suitable for mass-surveillance.
To the best of our knowledge, ours is the first work to attempt webpage fingerprinting attack for mass-scale surveillance. We intend to release our code and artefacts\footnote{Link to repository: https://gmit91@bitbucket.org/gmit91/snoopy.git}.
%%%%%%%%%%%%%%%
\begin{table*}[t]
\begin{centering}
\begin{tabular}{|l|c|c|c|c|c|c|}
\hline
\multicolumn{1}{|c|}{\multirow{3}{*}{\textbf{Existing works}}} & \multicolumn{6}{c|}{\textbf{Mass-surveillance Requirements}} \\ \cline{2-7} 
\multicolumn{1}{|c|}{} & \multicolumn{5}{c|}{Generalization Requirements} & \multirow{2}{*}{\begin{tabular}[c]{@{}c@{}}Compliance with a \\ finite query model\end{tabular}} \\ \cline{2-6}
\multicolumn{1}{|c|}{} & Caching & Cookies & \begin{tabular}[c]{@{}c@{}}Network \\ conditions\end{tabular} & \begin{tabular}[c]{@{}c@{}}User \\ interests\end{tabular} & \begin{tabular}[c]{@{}c@{}}Multi-tab \\ browsing\end{tabular} &  \\ \hline
\multicolumn{1}{|l|}{Cheng and Avnur~\cite{cheng1998traffic}, Sun et al.~\cite{sun2002statistical}} & {\xmark} & {\xmark} & {\cmark} & {\cmark} & {\xmark} & {\cmark} \\ \hline
\begin{tabular}[c]{@{}l@{}}Cai et al.~\cite{cai2012touching}, G. Danezis~\cite{danezis2010traffic}, Chapman and Evans~\cite{chapman2011automated},\\ Gong et al.~\cite{gong2012website}\end{tabular} & {\cmark} & {\xmark} & {\xmark} & {\xmark} & {\xmark} & {\cmark} \\ \hline
\multicolumn{1}{|l|}{Miller et al.~\cite{miller2014know}, Hayes and Danezis~\cite{hayes2016k}} & {\cmark} & {\cmark} & {\xmark} & {\xmark} & {\xmark} & {\cmark} \\ \hline
\begin{tabular}[c]{@{}l@{}}{Gu et al.~\cite{gu2015novel}, Zhuo et al.~\cite{zhuo2017website}}\end{tabular} & {\cmark} & {\cmark} & {\xmark} & {\cmark} & {\cmark} & {\xmark} \\ \hline
\begin{tabular}[c]{@{}l@{}}{Xu et al.~\cite{xu2018multi}}\end{tabular} & {\xmark} & {\xmark} & {\xmark} & {\xmark} & {\cmark} & {\xmark} \\ \hline
\multicolumn{1}{|l|}{Alan and Kaur~\cite{alan2019client}} & {\xmark} & {\cmark} & {\cmark} & {\xmark} & {\xmark} & {\cmark} \\ \hline
\multicolumn{1}{|l|}{Ghiette et al.~\cite{ghiette2020scaling}} & {\cmark} & {\cmark} & {\xmark} & {\cmark} & {\xmark} & {\xmark} \\ \hline
\multicolumn{1}{|l|}{Panchenko et al.~\cite{panchenko2016website}, Sirinam et al.~\cite{sirinam2019triplet}} & {\cmark} & {\cmark} & {\cmark} & {\cmark} & {\xmark} & {\xmark} \\ \hline
\multicolumn{1}{|l|}{Wang et al.~\cite{wang2021s}} & {\xmark} & {\cmark} & {\xmark} & {\cmark} & {\xmark} & {\cmark} \\ \hline
\multicolumn{1}{|l|}{Shen et al.~\cite{shen2020fine}} & {\xmark} & {\xmark} & {\xmark} & {\cmark} & {\xmark} & {\cmark} \\ \hline
\multicolumn{1}{|l|}{\textbf{Snoopy}} & {\cmark} & {\cmark} & {\cmark} & {\cmark} & {\cmark} & {\cmark} \\ \hline
\end{tabular}
\caption{Compliance of webpage fingerprinting techniques with the requirements for mass-surveillance}
\label{tab:literature}
\vspace{-10pt}
\end{centering}
\end{table*}

The rest of the paper is organized as follows. In Section~\ref{relwork} we present the literature related to our work and motivate the need for Snoopy. Next, we state our adversarial capabilities and assumptions, followed by a high level overview of Snoopy in Section~\ref{sec:framework}. Then we describe the design choices made for Snoopy in Section~\ref{sec:feature}. This is followed by a detailed description of Snoopy in Section~\ref{sec:details}. Thereafter, we describe the implementation and evaluation strategies of Snoopy and present a detailed analysis of the experimental results in Section~\ref{sec:evaluation}. Finally, we conclude the paper in Section~\ref{concl}.

%%%%%%%%%%%%%%%%%%%%%%%%%%%%%%%%%%%%%%%%%%%%%%%%%%%%%%%%%%%%%%%%%%%%%%%%%%%%%%%%%%%%%%%%%%
\section{Related Work}\label{relwork}
With growing security concerns, traffic on the Internet is encrypted. Thus, our work is related to the literature on Encrypted Traffic Analysis (ETA) techniques, which can be broadly classified into \textbf{(1)} Website identification attacks -- these works~\cite{liberatore2006inferring,panchenko2011website,gong2012website,wang2014effective,zhuo2017website,sirinam2019triplet,rahman2020tik} aim to infer coarse-grained web browsing information, i.e., the identity of websites browsed over Tor/VPN; and, \textbf{(2)} Webpage identification attacks -- these works~\cite{cheng1998traffic, sun2002statistical,danezis2010traffic,chapman2011automated,cai2012touching,miller2014know,gu2015novel,hayes2016k,panchenko2016website,xu2018multi,yan2018feature,alan2019client,ghiette2020scaling,shen2020fine,wang2021s} aim to infer fine-grained web browsing information of users from HTTPS traffic, i.e., the identity of webpages browsed within a website. 
The techniques used for webpage identification are very different from those used for website identification. On one hand, website identification require features (e.g., Round trip time) that vary across websites but remain consistent for webpages within a website. In contrast, webpage identification attacks~\cite{cheng1998traffic, sun2002statistical,danezis2010traffic,chapman2011automated,cai2012touching,miller2014know,gu2015novel,hayes2016k,panchenko2016website,xu2018multi,yan2018feature,alan2019client,ghiette2020scaling,shen2020fine,wang2021s} require features (e.g., HTML size) that vary across the webpages within a website, which can help distinguish between them. Blurring the lines between the two classes, some website identification attacks~\cite{liberatore2006inferring,panchenko2011website,wang2014effective,zhuo2017website,sirinam2019triplet,rahman2020tik} internally use webpage fingerprinting techniques to distinguish between the homepages of thousands of websites. We note that such techniques do not inherit the same set of challenges that a webpage identification attack faces (e.g., restrictions on number of accesses to the website). In this work, we focus on webpage identification attacks on a given website, subject to constraints on the number of webpage accesses.

In this section, we revisit the literature on webpage fingerprinting and assess their suitability for mass-scale HTTPS traffic analysis.
Webpage fingerprinting on mass-scale traffic requires a generalized classifier which could account for diversity in client behavior and network conditions. Diversity in client behavior, for example, in terms of browser/OS/device used, number of tabs open, and the sequence in which webpages are accessed, results in different fingerprints for the same webpage~\cite{alan2019client}. 
Diversity in network conditions in terms of jitter, bandwidth and packet drop rates also affects the webpage fingerprints~\cite{sirinam2019triplet}. Accounting for these diversities across so many factors is challenging due to practical restrictions on the number of website accesses. In this context, Table~\ref{tab:literature} classifies the prior works in terms of their generalization capabilities. %In the rest of this section we explore the evolution of webpage fingerprinting techniques in the literature of webpage identification attacks.
The earliest webpage fingerprinting techniques~\cite{cheng1998traffic, sun2002statistical} could successfully generalize webpage fingerprints for simple contemporary websites that did not use caching, cookies, and used static webpages with limited number of resources. Such works used basic features such as size of client request packets and sequences of encrypted object sizes, coupled with elementary statistical methods. 

The prevalence of web caching on the Internet prompted subsequent research works~\cite{danezis2010traffic, chapman2011automated,cai2012touching} to start the use of webpage-sequence fingerprinting to account for variations in traffic patterns based on the previously accessed webpage(s). 
However, the applicability of webpage-sequence fingerprinting techniques is restricted only to targeted attacks and are not suitable for mass-scale surveillance due to the following reasons -- \textbf{(1)} It necessitates an enormous number of website accesses as compared to individual webpage based fingerprinting techniques, even for medium-sized websites;
and, \textbf{(2)} these works used features such as resource ordering~\cite{danezis2010traffic} and traffic burst patterns~\cite{chapman2011automated,cai2012touching} for fingerprinting webpage sequences. Such features are not consistent across varying network conditions and user behavior (e.g. number of browser tabs open), and hence, require an estimate of the network speed and browsing behavior of the victim.  
Such fingerprinting techniques were useful in the context of surveillance on a small set of targeted users, since the attackers had knowledge about the targeted victim's browsing behavior and network conditions. However, assuming interests of website users, their browsing behavior (sequential or single-tab browsing) and their network conditions is impractical for mass-surveillance in realistic scenarios~\cite{juarez2014critical}.

The use of cookies by the next generation websites further complicated the process of webpage fingerprinting, even for targeted attacks. Tracking cookies embedded inside web resources and session cookies included in application layer headers result in resource size variations. To account for these variations, subsequent research works~\cite{miller2014know,hayes2016k} either used complex techniques (combination of clustering algorithms, Gaussian distribution, Hidden Markov Model and Viterbi algorithm) complemented by huge datasets~\cite{miller2014know} or simplified their assumptions about user behavior~\cite{hayes2016k}. The former technique~\cite{miller2014know} is too restrictive about user interests since they consider only a small set of webpages to make data collection practical. On the other hand, the latter one~\cite{hayes2016k} assumes impractical user behavior such as single tab browsing without caching. \hlt{However, these techniques are restricted only to targeted attacks, and are not suitable for mass-scale surveillance due to the following reasons -- \textbf{(1)} The goal of mass-surveillance is to understand the interests of the users of a website. Therefore, restrictive assumptions about user interests are unreasonable in the context of mass-surveillance, and; \textbf{(2)} Restrictive assumptions regarding the browsing behavior (for e.g., the number of tabs used, OS/Browser used, and the browser configuration) of a diverse set of website users are also unreasonable.}
Therefore, these techniques cannot be used for analysis of encrypted traffic on a mass scale.

Recent works~\cite{gu2015novel,panchenko2016website,zhuo2017website,xu2018multi,sirinam2019triplet,alan2019client,shen2020fine,ghiette2020scaling,wang2021s} on webpage fingerprinting have recognized the importance of practical mass-scale surveillance. Most of these works acknowledge the necessity of accounting for diversity in user interests~\cite{panchenko2016website,ghiette2020scaling,wang2021s}, user behavior (particularly, a wide variety of user interests and multi-tab browsing)~\cite{gu2015novel,zhuo2017website,xu2018multi}, OS/browser settings~\cite{alan2019client} and network conditions~\cite{sirinam2019triplet} while performing encrypted traffic analysis on mass-scale web traffic. However, it is to be noted that each of these works can only generalize for at most one of these factors. This is because, for most of these works~\cite{zhuo2017website,alan2019client,sirinam2019triplet}, generalizing for even one of these factors using the features and/or the techniques presented in these works requires collecting a massive amount of traffic samples, coupled with a heavyweight ML algorithm. This would result in a \hlt{large number of website queries (i.e., the number of website accesses done by the adversary)}, as well as a high bootstrap time every time the website contents change or a new OS or browser or firmware becomes popular. 
For instance, one of these works~\cite{alan2019client} account for diversity in browser, OS and device used by the client at the cost of an immensely high bootstrap time. 
% {\color{red}Also, it is not possible to generalize across multi-tab browsing scenarios even with a finitely large number of samples, since in practice, a user can open and close a new tab at any random point of time within the browsing session.}
%On the other hand, some of these recent works~\cite{shen2020fine} focus on reducing the time required to collect network traces and build the prediction model (i.e., the bootstrap time) every time the contents of a website change. This would ensure timeliness of the attack and enhance its effectiveness in practical scenarios. However, it achieves that at the cost of assuming background knowledge about the browser, OS and device used by the client, which is unrealistic in case of mass-scale surveillance. 

\hlt{As summarized in Table~\ref{tab:literature}, existing ETA techniques have complied with a finite query model at the cost of restrictive assumptions about the interests and browsing behavior of the users, and vice-versa, and are suitable for mass-scale surveillance. We propose \textit{Snoopy}, an ETA technique that is designed with the goal of meeting the generalization requirements and complies with a finite query model, both of which are essential for mass-surveillance.}

% By analyzing the existing techniques (as shown in Table~\ref{tab:literature}), we conclude that, compliance with a finite query model has so far come at the cost of restrictive assumptions about the interests and browsing behavior of the users, and vice-versa. %Similarly, accounting for variations in the interests and browsing behavior of the users or in caching and cookie settings makes the solution non-compliant with a finite query model. 
% Therefore, none of the existing works are suitable for mass-scale surveillance. We propose \textit{Snoopy}, an ETA technique that is designed with the goal of meeting the generalization requirements and complies with a finite query model, both of which are essential for mass-surveillance.
%On the other hand, Snoopy, our proposed mass-scale webpage fingerprinting technique achieves all the generalization requirements, yet complies with a finite query model.
%%%%%%%%%%%%%%%%%%%%%%%%%%%%%%%%%%%%%%%%%%%%%%%%%%%%%%%%%%%%%%
%%%%%%%%%%%%%%%%%%%%%%%%%%%%%%%%%%%%%%%%%%%%%%%%%%%%%%%%%%%%%%
\section{The Snoopy Framework}\label{sec:framework}
In this section, we present \textit{Snoopy}, our proposed adversarial framework for mass-surveillance through webpage fingerprinting
% analysis of mass-scale encrypted web traffic for webpage identification}.
We first describe the capabilities of the adversary, then define the scope of our work, and finally provide a high level overview of Snoopy.

\subsection{Adversary Capabilities}
Our adversary is a compromised network device on the client-server path that can \textbf{(1)} access unencrypted header fields of both control and data packets; and, \textbf{(2)} observe the traffic characteristics, such as the size of encrypted packets.
% ; and, \textbf{(3)} delay packets, limit the bandwidth of the transit link between the client and the server, and drop packets. 
Such an adversary model is not only realistic but also common today. In the real world, this translates to a rogue Internet Service Provider (ISP) or a malicious entity who compromises an ISP router.

% All Internet Service Providers (ISPs) or entities with similar level of control and visibility have the capability to monitor the encrypted traffic traces of website visitors~\cite{shodan}.
% At the outset, such an adversary might seem stronger than the basic passive adversary model which would only capture network traffic in promiscuous mode. However, in reality, the privilege required (i.e., superuser access) by the basic adversary model is sufficient for tuning network parameters such as delay and bandwidth. Such an adversary model is not only realistic but also common today~\cite{shodan}. 

\subsection{Assumptions and Scope}
We make the following assumptions about the context in which Snoopy is to be employed.
\begin{itemize}[leftmargin=*]
\item We do not consider websites that host dynamic or highly personalized content for each user, for e.g., social media websites and web search engines, as targets for our attack; 
% Websites undergoing A/B testing are also out of scope of this paper;
\item The adversary does not have the capability to decrypt the traffic of real website users. Most real-world adversaries cannot decrypt web traffic except maybe those involving government agencies or authorized middle-boxes~\cite{authorized-surveillance}; and,
\item Snoopy must compromise a network device that has access to all the encrypted Application layer packets (referred to as \textit{traffic trace}) exchanged between the client and the server for the entire duration of a browsing session. For instance, our framework cannot be used when route flapping (dynamic change of route due to, for instance, unreliable connections) occurs on the server-client path. 
\end{itemize}

\subsection{High Level Overview}
\begin{figure}[t]
    \centering
    \includegraphics[scale=0.105]{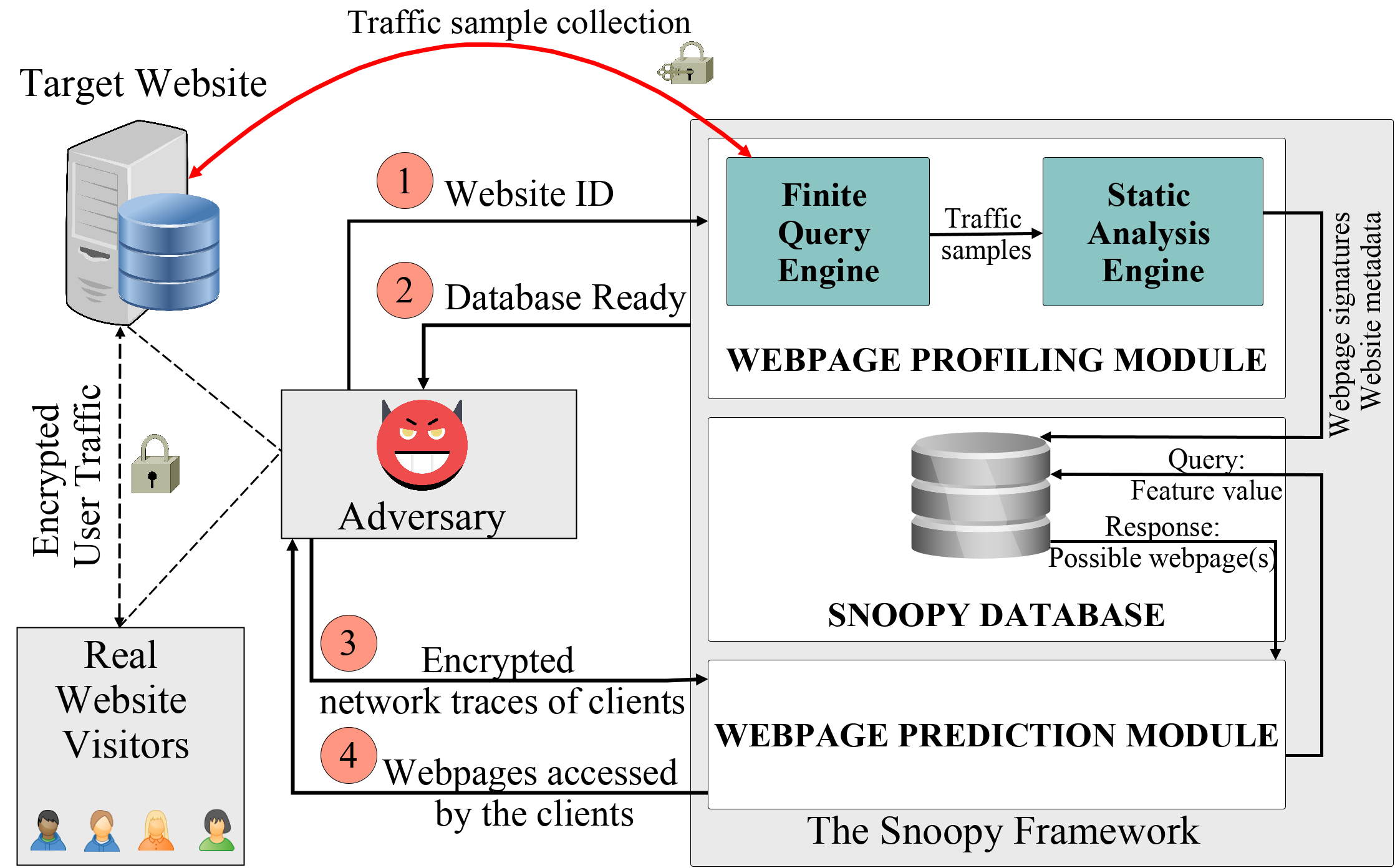}
    % \caption{A High Level Overview of Snoopy\vspace{-15pt}}
    \caption{The components of Snoopy}
    \label{fig:snoopy-components}
    % \vspace{-15pt}
\end{figure} 
We now provide a high level overview of Snoopy -- our mass-surveillance framework. Snoopy comprises the \emph{Snoopy Database} and two functional modules, namely, \emph{Webpage Profiling Module} and \emph{Webpage Prediction Module}. The Snoopy Database is meant to store information (i.e., webpage fingerprints and additional website metadata) required for carrying out the predictions, and the two functional modules are used by the adversary for profiling webpages and predicting the identities of webpages accessed by the users. Figure~\ref{fig:snoopy-components} shows the different components of Snoopy and how an adversary interacts with them. 

To populate the Snoopy Database, the adversary first triggers the Webpage Profiling Module of Snoopy with the Website ID (IP address or homepage URL) of the target website, as shown in Step 1 of Figure~\ref{fig:snoopy-components}. 
The goal of this module is to gather information about the website that will be useful to the Webpage Prediction Module. The input to this module is the website identifier (IP address or homepage URL) and the set of features to be used for webpage fingerprinting. First, the finite query engine in this module performs a focused traffic sample collection from the target website. Subsequently, this traffic is decrypted and both the encrypted and decrypted versions of the traffic samples are passed on to the static analysis engine. While Snoopy does not require the decryption keys of users during the attack, it still needs to decrypt the traffic samples that it generates on its own during the profiling stage. This is necessary to identify the plaintext resources (e.g., HTML, javascript, images) of each webpage from their encrypted counterparts. The final output of the Webpage Profiling module includes information pertaining to \textbf{(1)} the structure of the website; \textbf{(2)} resource download sequences for each webpage in the website; \textbf{(3)} signature of these web resources; and, \textbf{(4)} other relevant website metadata, such as cache-ability and cookie information of the web resources. 

Once the database has been populated, Snoopy notifies the adversary that it is ready to be used for prediction, as shown in Step 2 of Figure~\ref{fig:snoopy-components}. When the adversary wants to predict the webpages accessed by a user from their encrypted traffic trace, it triggers the \emph{Webpage Prediction Module} of Snoopy. This module takes an encrypted traffic trace $T$ of a user as input, as shown in Step 3 of Figure~\ref{fig:snoopy-components}, and predicts the webpages that are accessed in $T$ using the information stored in the Snoopy database. Snoopy performs this prediction in two steps. First, it extracts the feature values from the input trace $T$ and performs a lookup on the Snoopy Database to retrieve the sequence of candidate web resources present in $T$. Next, Snoopy uses this sequence and the website metadata to predict the final set of webpages, which it returns to the adversary (Step 4 of Figure~\ref{fig:snoopy-components}). 
% Mention what exactly it matches

% Figure~\ref{fig:snoopy-hlov} shows users accessing a targeted website over an encrypted channel in the presence of Snoopy, our proposed on-path adversary. When the targeted website and the encrypted network trace of the user are given as inputs, Snoopy outputs the possible webpage(s) accessed by the user. Contrary to most works in the literature, Snoopy can predict user activities irrespective of the browsing context. For instance, Snoopy would work irrespective of the OS/Browser used, the number of webpages accessed, the sequence in which webpages are accessed, number of tabs opened by the user, and the network condition of the user. Additionally, Snoopy can also work across different application layer protocols (HTTP/1.x, HTTP/2, HTTP/3) and across websites of different sizes. 
% When the user from Figure~\ref{fig:snoopy-hlov} requests the server for a webpage, the server transmits a sequence of encrypted web resources (an HTML file followed by images, Javascript files, style files, video files, etc.) as a response. The resultant traffic pattern, although encrypted, helps Snoopy predict the webpage requested by the user. Snoopy first fingerprints all possible webpages in the website by accessing them individually. The fingerprint of each webpage is stored in a database. Later Snoopy matches the user's traffic pattern that it sees at a compromised router with the pre-computed fingerprints in the database to determine the webpage(s) accessed. 

The most unique and crucial feature of Snoopy is its ability to comply with a finite query model, while retaining its generalization capability at the same time. Achieving this balance between generalization and a finite number of queries is extremely challenging. 
% Our studies presented in Section~\ref{sec:intro} and Section~\ref{relwork} show how existing works often compromise any one of these to achieve the other. Existing works that focus on generalization~\cite{} use complex traffic features and a large number of traffic samples to model the variations in feature values across different browsing contexts. Snoopy follows an alternative approach for solving this problem. 
Snoopy attempts to solve this problem using the following:
\begin{description}
    \item[(1) Predictable features: ] It uses a feature that exhibits negligible or predictable variation across browsing contexts. The predictability of the feature values allows Snoopy to fingerprint webpages in different browsing contexts even with a finite number of traffic samples, complying with a finite query model. The feature used in Snoopy and the justification behind this choice are discussed in further details in Section~\ref{sec:feature}; and,
    \item[(2) Focused data collection and feature value estimation:] For collecting traffic samples from the targeted website, Snoopy uses a focused data collection method to account for certain browsing contexts, which has been discussed in details in Section~\ref{sec:profiling}. Snoopy estimates the variations in feature values in different browsing contexts using static analysis of HTTP headers and payload. At the time of prediction, Snoopy uses these estimated feature values (from the Snoopy Database) based on the user's browsing context. The feature value estimation and prediction techniques used by Snoopy are detailed in Section~\ref{sec:analysis}. 
\end{description}
%\textbf{(1)} using a feature that exhibits negligible or predictable variation across browsing contexts. The predictability of the feature values allows Snoopy to fingerprint webpages in different browsing contexts even with a finite number of traffic samples, complying with a finite query model. The feature used in Snoopy and the justification behind this choice are discussed in further details in Section~\ref{sec:feature}. For collecting traffic samples from the targeted website, Snoopy uses a focused data collection method to account for certain browsing contexts, which has been discussed in details in Section~\ref{sec:profiling}; and, \textbf{(2)} estimating the variations in feature values in different browsing contexts using static analysis of HTTP headers and payload. At the time of prediction, Snoopy uses these pre-computed feature values (from the Snoopy Database) based on the user's browsing context. The prediction technique used by Snoopy is detailed in Section~\ref{sec:prediction}. 
% In the subsequent sections, we explore how such a simple technique helps to strike a balance between generalization and the constraints imposed by a finite query model. 
% The feature used in Snoopy and the justification behind this choice are discussed in further details in Section~\ref{sec:feature}. The Webpage Profiling Module, that carries out focused data collection and static analysis are discussed in details in Section~\ref{sec:profiling}. Finally, the prediction technique is detailed in Section~\ref{sec:prediction}. 

\section{Snoopy Design: Predictability of Fingerprints}\label{sec:feature}
Snoopy needs an encrypted traffic feature that either exhibits no variation or predictable variation across different browsing contexts, in order to minimize the number of traffic samples required during webpage profiling. In other words, the traffic features used for webpage profiling need to be \textit{stable} across different browsing contexts. For this, we evaluated different classes of traffic features (e.g., timing side-channels and traffic burst patterns) that are widely used in prior ML-based works~\cite{gong2012website,miller2014know} for targeted attacks and assessed their stability. It is known and also shown in our evaluations presented in the Appendix that such traffic features have poor stability and are therefore not predictable. For instance, the traffic burst pattern corresponding to a webpage download changes with variations in network conditions and the number of parallel browser tabs used by a user.

% Snoopy primarily uses \textit{encrypted web resource sizes}, which is computed as the sum of TLS segment sizes of all packets carrying the resource, to fingerprint.
Snoopy uses \textit{sequence of encrypted web resource sizes} as the fingerprint of a webpage. The size of an encrypted resource is computed as the sum of TLS segment sizes of all packets carrying the resource.
Since each user might have a different browsing context, Snoopy first focuses on {\em understanding} the effect of various browsing contexts on the fingerprint. While some of the factors of the browsing context affect the encrypted resource size, some others affect the sequence in which the resources are downloaded. 
% Understanding the effect of these factors on the sequence of web resource sizes helps Snoopy to compute the feature values for one browsing context from the feature values corresponding to a different browsing context, thereby eliminating the need for collecting traffic samples for all possible browsing contexts.
\hltr{An analysis of these factors helps Snoopy \textit{statically} estimate the variations caused by each of these factors, thereby eliminating the need for collecting traffic samples for all possible browsing contexts.}
We now explain how the size of encrypted resources and their download sequence change based on the browsing context of a user: 

\begin{description}[leftmargin=0.2cm]
\item[\textbf{(1)} \textit{Operating system (OS) --}] The choice of OS affects the TLS segment size of packets in multiple ways. First, the  TLS segment size of a web resource depends on the TLS implementation on the corresponding OS. Different OS'es start with the same TLS record but they break them into segments of different sizes. Although one might not expect this to affect the TLS segment size, we noticed minor differences when the sum of TLS segment sizes is calculated. A deeper analysis revealed that the number and size of segments affects the metadata associated with every TLS segment in a number of ways, which are, to our advantage, predictable. First, there are some TLS headers added to every TLS segment. Therefore, if a record is broken into several small TLS segments, the total number of TLS header bytes for all the segments would be greater than the total number of TLS header bytes added in case the record was broken into fewer larger segments.
% the MAC field and the length field values vary in length depending on the size of the segment, thereby affecting the TLS segment sizes.
Second, a variable field on the HTTP header, namely, the user agent string that carries the name of the OS also affects the TLS record size.
\item[\textbf{(2)} \textit{Browser --}] The browser name  indirectly affects the TLS segment size since it is also a part of the User Agent string. As different browser names have different lengths, they affect the length of the User Agent string, which in turn, results in variation in the TLS record sizes.

\item[\textbf{(3)} \textit{Browsing sequence --}] The sequence in which a user browses the webpages affects both the TLS record sizes and the download sequence of the web resources, particularly when caching and cookies are allowed (enabled in most browsers by default). 

\SubItem {\textbf{Caching:}} When a user visits a webpage containing a resource that was previously downloaded, the resource may not get downloaded if resource caching is enabled.
This results in variation in resource download sequence. For instance, if a user visits webpage $W_X$ (composed of resources $r_1$ and $r_2$), followed by webpage $W_Y$ (composed of resources $r_1$ and $r_3$), the overall resource download sequence would be $r_1$--$r_2$--$r_3$. On the other hand, if caching was disabled, it would result in the following sequence: $r_1$--$r_2$--$r_1$--$r_3$.

\SubItem {\textbf{Cookies:}} When a user allows a website to use cookies, the server sends a \textit{session cookie} in the HTTP header along with the first resource delivered during the browsing session. Addition of the session cookie increases the TLS record size of the first resource downloaded during a browsing session. Therefore, a resource would have a larger TLS record size if it gets downloaded at the beginning of a browsing session, as compared to its TLS record size when it gets downloaded at a later point during the browsing session. Furthermore, another type of cookie, called the \textit{tracking cookie}, affects the payload size of the transmitted resource. Tracking cookies hold information about the browsing behavior of a user such as the URL of the previously browsed webpage(s) in the session. Since different webpage URLs have different lengths, the variation in TLS record size due to tracking cookies depend on the webpage(s) last visited by a user. Also, tracking cookies hold a null value for the first resource delivered during the browsing session. Again, this causes the TLS record size of a resource to vary depending on the user's browsing sequence. For instance, if the size of a resource $r_1$ is $s_1$ and the size of the session cookie is $sc$, then the TLS record size of $r_1$ would be $(s_1 + sc)$ if $r_1$ is the first resource to be downloaded. On the other hand, if $r_1$ is downloaded after the user has visited a webpage of URL length $tc$ (also, the tracking cookie size) in a browsing session, the TLS record size of $r_1$ would be $(s_1 + tc)$.  

\item[\textbf{(4)} \textit{Application layer protocol --}] Grouping packets that carry a resource is critical for computing the encrypted resource size. For HTTP/1.x, packets belonging to a resource have the same TCP ACK number, making the process straightforward. However, it is not the same for HTTP/2 and HTTP/3 websites due to pipelining and multi-threaded server operations. With multi-threading, packets belonging to two different resources could get interleaved within the same TCP stream (in HTTP/2~\cite{http2}) or QUIC stream (in HTTP/3~\cite{http3}). To handle such complex scenarios, Snoopy adopts the technique described in our recent work~\cite{mitra2020depending} for computing encrypted resource sizes in HTTP/2 websites. For HTTP/3 websites, Snoopy drops the QUIC connection establishment packets so that the communication protocol falls back to HTTP/1.1 or HTTP/2.
\item[ \textbf{(5)} \textit{Parallel tabs --}] Browsing on concurrent tabs affects the sequence of downloaded resources. For instance, when a user browses two webpages, say $W_X$ and $W_Y$ one after another on two tabs, the resources of $W_X$ get downloaded first, followed by the resources of $W_Y$. On the other hand, if the user opens the two webpages in two parallel browser tabs, the resources of $W_X$ and $W_Y$ would get downloaded in an interleaved fashion. In such cases, the attacker faces the additional challenge of identifying the resources download sequence corresponding to each webpage from the interleaved sequence of encrypted resources, which makes the webpage prediction process more challenging.  
\item[\textbf{(6)} \textit{Network conditions --}] %Encrypted resource sizes can be affected by poor network conditions. 
A congested link results in packet transmission delays, packet drops, and blocking of new connections. Among these, packet drops affect the encrypted size of resources. The dropped packets may or may not get re-transmitted, based on the nature of the resource. 
% In the event of packet re-transmissions, the re-transmitted packets need to be assembled properly for Snoopy to work. This action can be easily performed using state-of-the-art network protocol dissectors. 
In the event of packet re-transmissions, the re-transmitted packets can be easily assembled using state-of-the-art network protocol dissectors so that Snoopy can still work. 
Therefore, computation of the encrypted resource size by Snoopy is not affected in this case. However, in rare situations when a large number of dropped packets are not re-transmitted or results in a broken connection, Snoopy cannot compute or predict the resource size correctly. 
Severe network congestion may also result in route flapping. In that case, the attacker would not be able to access the complete network trace unless it has control over the new route too. Such circumstances cannot be handled by Snoopy.  
\end{description}
 
% The main objective of Snoopy is to achieve generalization across browsing contexts. 
% However, this becomes extremely challenging when we need to comply with a limited query model, wherein the adversary is allowed only a limited number of attempts to sample a given website.
\hlt{Presently, Snoopy has been configured to use \textit{sequence of encrypted web resource sizes} as the feature for fingerprinting webpages}. We incorporate the aforementioned knowledge in the design of Snoopy to make it compliant with a limited query model while allowing it to be generalized at the same time. However, if a feature that is more stable is discovered, Snoopy can be configured to use that instead. 
% Snoopy achieves the combination of generalization and compliance to limited query model by \textbf{(1)} following a focused data collection method that evades most state-of-the-art DDoS detection techniques and assures a low bootstrap time; and, \textbf{(2)} employing static analysis techniques to pre-compute the deterministic variations in feature values due to variations in the browsing contexts. The data collection and static analysis methods are discussed next in Section~\ref{}.

%%%%%%%%%%%%%%%%%%%%%%%%%%%%%%%%%%%%%%%%%%%%%%%%%%%%%%%%%%%%%%%%%%%%%%%%%%%%%%%%%%%%%%%%%%%%%%%%%%%%%%%%%%%%%%%%%%%%%%%%%%%%%%%%%%%%
\section{Snoopy Design: A Deep Dive}\label{sec:details}
In this section, we provide a detailed description of the two functional modules of Snoopy, viz., the Webpage Profiling Module and the Webpage Prediction Module, and how they use the insights from Section~\ref{sec:feature}. As discussed in Section~\ref{sec:feature}, the present implementation of Snoopy uses only one feature -- sequence of encrypted resource sizes. Therefore, in the rest of this section, we describe the functional modules of Snoopy in the context of this feature.

\subsection{Webpage Profiling Module}\label{sec:profiling}
The input to this module is the target website identifier $WebsiteIP$, the encryption function $EF$, and the webpage fingerprinting feature $F$ (i.e., encrypted resource size sequence). The output of the webpage profiling module includes the following information \textbf{(1)} the \textit{structure of the website}, represented by a graph $\mathbb{G}$ = $(\mathbb{W}, E)$, where, $\mathbb{W} = \{w_1, w_2, \ldots, w_n\}$ is the set of $n$ distinct webpages in the website and  $E$ = \{$(w_i, w_j)$ | $w_j$ \textit{is directly navigable from} $w_i$\} is the set of directed edges; \textbf{(2)} a \textit{resource-map} $\mathbb{RM} : \mathbb{W} \longrightarrow \mathbb{R}$ that shows the relationship between webpages in $\mathbb{W}$ and the web resources in $\mathbb{R}$, where $\mathbb{R} = \{r_1, r_2, \ldots, r_m\}$ is the set of web resources that constitute the targeted website ; \textbf{(3)} the \textit{set of resource download sequences} for all $n$ webpages in the website, denoted by $\mathbb{S} = \{S_1, S_2, \ldots, S_n\}$, where $S_i$ be the sequence of resources requested by the client when a user accesses the webpage $w_i$; \textbf{(4)} \textit{signatures of web resources}, i.e., the encrypted size of each resource; and, \textbf{(5)} other \textit{relevant website metadata}, such as the cache-ability and cookie information of the web resources. 
% Once the database has been populated, Snoopy notifies the adversary that it is ready to be used for prediction, as shown in Step 2 of Figure~\ref{fig:snoopy-components}. 
We now describe the working of the Webpage Profiling Module using Algorithm~\ref{alg:1} in the following sections.

\subsubsection{Extracting Structure of the Website}
\begin{figure}[t]
    \centering
    \includegraphics[scale=0.21]{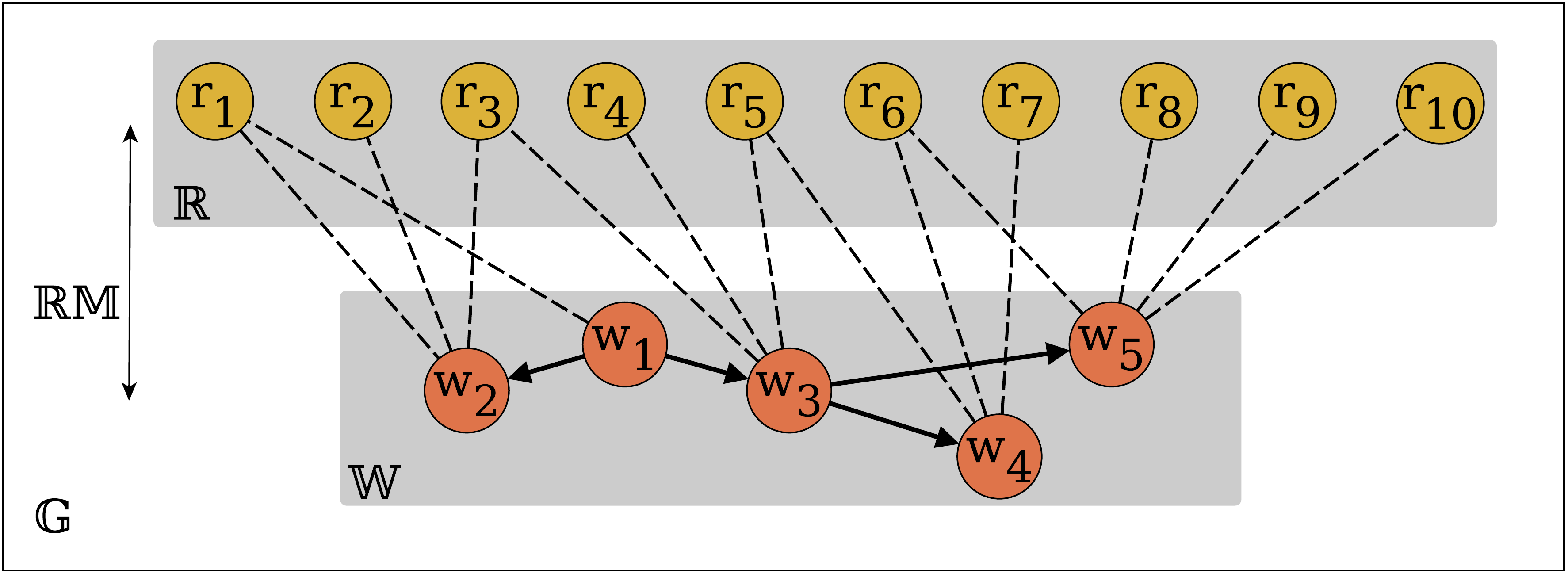}
    \caption{Representation of the structure of a sample website}
    \label{fig:resmap}
    % \vspace{-15pt}
\end{figure}

The GET\_WEBSITE procedure in Step~\circled{1} in Algorithm~\ref{alg:1} uses automated website-crawling and webpage parsing techniques to construct $\mathbb{G}$, the graph representation of the website of interest. A vertex $u$ in $\mathbb{G}$ has a directed edge to another vertex $v$ if $u$ represents a webpage that has embedded hyperlinks to help users navigate directly to the webpage represented by $v$.   
Following this, the GET\_RESOURCES procedure in Step~\circled{2} parses each webpage in $\mathbb{W}$ to extract its resource download sequence and compiles $\mathbb{R}$, the list of embedded web resources present in the website. The set of resource download sequences thus retrieved is denoted by $\mathbb{S}$, and the unique set of embedded resources is denoted by $\mathbb{R}$.
The MAP\_RESOURCES procedure in Step~\circled{3} creates a bipartite graph $\mathbb{RM}$ where the vertices consist of elements in $\mathbb{W}$ and $\mathbb{R}$. A webpage $w_i$ is mapped to a web resource $r_j$ when $r_j$ is embedded in $w_i$, and this mapping is denoted by an edge between the vertices corresponding to $w_i$ and $r_j$.
Figure~\ref{fig:resmap} shows an example of $\mathbb{R}$, $\mathbb{W}$, $\mathbb{RM}$, and $\mathbb{G}$ for a website. 
\subsubsection{Focused Traffic Sample Collection and Formulating Signatures of Web Resources}\label{sec:sample-collection}
For building webpage fingerprints Snoopy requires a dataset comprising traffic samples that capture how different web resources manifest themselves on an encrypted communication channel in different browsing contexts. 
The COLLECT\_SAMPLES procedure in Step~\circled{4} of Algorithm~\ref{alg:1} builds this dataset. The inputs to this step are the set of webpages $\mathbb{W}$ and the fingerprinting feature $F$, i.e., sequence of encrypted resource sizes. As discussed in Section~\ref{sec:feature}, the values of this feature vary with different browsing contexts. We now discuss how the COLLECT\_SAMPLES procedure handles these different scenarios. 
\begin{itemize}[leftmargin=.2in]
    \item [\textbf{(1)}] \textbf{Factors causing no variation in feature values.} As discussed in Section~\ref{sec:feature}, the values of encrypted resource sizes remain invariant to changes in factors such as network conditions. Snoopy does not require traffic samples to account for such factors that do not affect the feature values. 
    \item [\textbf{(2)}] \textbf{Factors causing predictable variation in feature values.} As discussed in Section~\ref{sec:feature}, the variations in encrypted resource sizes due to factors such as operating system, browser, and number of parallel tabs used are deterministic in nature. 
    % For such factors, Snoopy does not require traffic samples from all possible browsing contexts.
    Differences in feature values due to these factors can be estimated by Snoopy from the domain knowledge of network and browser protocols incorporated in its design. Therefore, no extra traffic samples need to be collected for accounting for such browsing contexts. The COLLECT\_SAMPLES procedure collects traffic samples for any one browsing context (for example, Firefox browser). Snoopy can estimate the feature values for the other browsing contexts (for example, Google Chrome browser) using static analysis techniques described in Section~\ref{sec:analysis}.
    \item [\textbf{(3)}] \textbf{Factors causing website-specific variation in feature values.} Snoopy requires traffic samples to estimate feature values due to variations in factors that depend on website design -- the resources that can be cached and the information that the cookie carries. To handle this, the COLLECT\_SAMPLES procedure collects traffic samples from each webpage once with caching on and cookies allowed and once with caching off and cookies prohibited.
\end{itemize}

To collect the traffic samples for estimating the feature values, the COLLECT\_SAMPLES procedure spawns multiple dummy clients to access the webpages, one at a time, in separate sessions. During each webpage access, the encrypted traffic of each client is captured by a network monitoring tool from the beginning till the end of the browsing session.
% These encrypted traces constitute the dataset. 
Subsequently, Snoopy processes every traffic trace and splits them into multiple sub-traces, corresponding to each resource $r_i$. Note that the packets carrying the same resource can be easily identified from an encrypted trace, since they have the same TCP acknowledgement number. The actual resource that is carried by the packets in a sub-trace are found by decrypting the sub-trace with the Transport Layer Security (TLS) keys used by the dummy clients. The sub-traces and the resources they correspond to are stored in the dataset named $TSamples$, where each entry is of the form \{$\text{\em{sub-trace}}$, $r$\}.
% This process is repeated $N$ times, under different browsing conditions to broaden the applicability of the model.
Once the $TSamples$ dataset is built, the feature values of the resources can be extracted for building the resource signatures.
\begin{algorithm}[t]
        {%\footnotesize
        \footnotesize 
        \caption{PROFILE\_WEBSITE}\label{alg:1}
        \setcounter{AlgoLine}{0}
        \SetAlgoLined
        \SetKwInOut{Input}{input}
        \SetKwInOut{Output}{output}
        \Input{$WebsiteIP$, $EF$, $F$}
        \Output{$\mathbb{G}$, $\mathbb{S}$, $\mathbb{RM}$,  $reverseFeatureDB$, $cookie\_var$}
        %\vspace{1mm}
        \tcc{automated website crawling}
        $\mathbb{G}$ = GET\_WEBSITE($WebsiteIP$)\;
        \tcc{parsing hypertext}
        $\mathbb{S}$, $\mathbb{R}$ = GET\_RESOURCES($\mathbb{W}$)\;
        \tcc{webpage-resource relation}
        $\mathbb{RM}$ = MAP\_RESOURCES($\mathbb{W}$,$\mathbb{R}$)\;
        \tcc{traffic sampling for webpage access}
        $TSamples$ = COLLECT\_SAMPLES($\mathbb{W}$, $F$)\;
        \tcc{unique identifiers for encrypted resources}
        $featureDB$ = BUILD\_SIGNATURES($F$, $\mathbb{R}$)\;
        \tcc{look-up table for prediction}
        $reverseFeatureDB$ = CONSTRUCT\_DICTIONARY($featureDB$)\;
        \tcc{variable cookie fields}
        $cookie\_var$ = COMPUTE\_COOKIE\_VAR($\mathbb{G}$, $featureDB$, $EF$)
}
\end{algorithm}

\subsubsection{Constructing Feature and Resource Databases}\label{sec:profiling-databases}
% The feature $F$ that we use in this paper for resource identification is the {\em sequence of encrypted resource size} (TLS record size of the encrypted resources). 
% As we will subsequently see, this feature exhibits a high-level of stability -- i.e., the value of this feature deviates the least due to variations in browsing context of the users as well as network conditions. In fact, this feature is also highly stable when compared to features such as round trip download time and packet size sequence, which have been used in prior works. Snoopy has been configured to use this as a side-channel by default. However, if a side-channel that is more stable is discovered, Snoopy can be configured to use it instead. 

In Step~\circled{5} of Algorithm~\ref{alg:1}, the BUILD\_SIGNATURES procedure extracts the $F$ value (encrypted size) of each resource from its corresponding sub-trace, which essentially captures the effect of the encryption function $EF$. \hlt{For instance, if a sub-trace contains two packets of size $50$ bytes each, the $F$ value of the corresponding resource would be $100$ bytes.}
For every extracted sub-trace in $TSamples$, an entry of the form $<r_i, sig_i>$ is made in the signature database $featureDB$, where $sig_i$ is the $F$ value of the resource $r_i$. 

During prediction, an inverse mapping (i.e., $EF^{-1}$) from an observed feature value ($f_i$) to the corresponding resource ($r_i$) will be required. However, constructing the inverse mapping $EF^{-1}$ is not straightforward.

\noindent\textbf{Challenges in constructing $EF^{-1}$.} Constructing $EF^{-1}$ is challenging because it is not possible to perform a one-to-one mapping between encrypted resource sizes and the corresponding resources. This is because, as we discussed in Section~\ref{sec:feature}, the sizes of encrypted resources vary due to variations in user browsing contexts. Therefore, a single resource might have different sizes in different browsing contexts. Likewise, a given size value might correspond to multiple resources in a given browsing context. Moreover, due to the finite query model, Snoopy cannot collect traffic samples (data points) for all possible browsing contexts. 

%Therefore, instead of performing a one-to-one mapping between resource sizes and resources, the Website Profiling Module performs a many-to-many mapping between encrypted resource sizes and resources. Additionally, this module also computes and stores some extra meta information about the resources with respect to different browsing contexts (for e.g., user agent string length, cache-ability of resources, cookie values, etc.). During prediction, Snoopy uses these meta information to estimate the encrypted resource sizes for a given browsing context. 

%The CONSTRUCT\_DICTIONARY procedure performs the many-to-many mapping between encrypted resource sizes and resources, and builds a reverse database $reverseFeatureDB$ out of $featureDB$ in Step~\circled{6}. 
The CONSTRUCT\_DICTIONARY procedure builds a reverse database $reverseFeatureDB$ out of $featureDB$ by taking into account the many-to-many mapping between encrypted resource sizes and resources. 
Each entry in $reverseFeatureDB$ is of the form $<f_i, L_i>$, where $f_i$ corresponds to a unique $FS$ value and $L_i$ is a list of resources that have this value. Note that $L_i$ may include (1) multiple resources that have the same signature and (2) multiple instances of the same resource (each corresponding to a dummy client access). Figure~\ref{fig:featuredb} shows an instance of featureDB and reverseFeatureDB constructed from the $TSamples$ collected corresponding to a website.

During prediction, to narrow down on the possible resources corresponding to a sub-trace, Snoopy uses the meta information (i.e., user-agent string, cache-ability, and cookies) to estimate the possible encrypted resource sizes in a given browsing context. 
Although the estimations for cache-ability and user-agent string can be determined easily (refer Section~\ref{sec:sample-collection}), for cookies, we need to collect additional traffic samples. The COMPUTE\_COOKIE\_VAR procedure (Step~\circled{7} of Algorithm~\ref{alg:1}) processes these traffic samples to extract and store the meta information pertaining to cookies as follows: 

\begin{figure}[t]
	\vspace{2pt}
    \centering
    \includegraphics[scale=0.32]{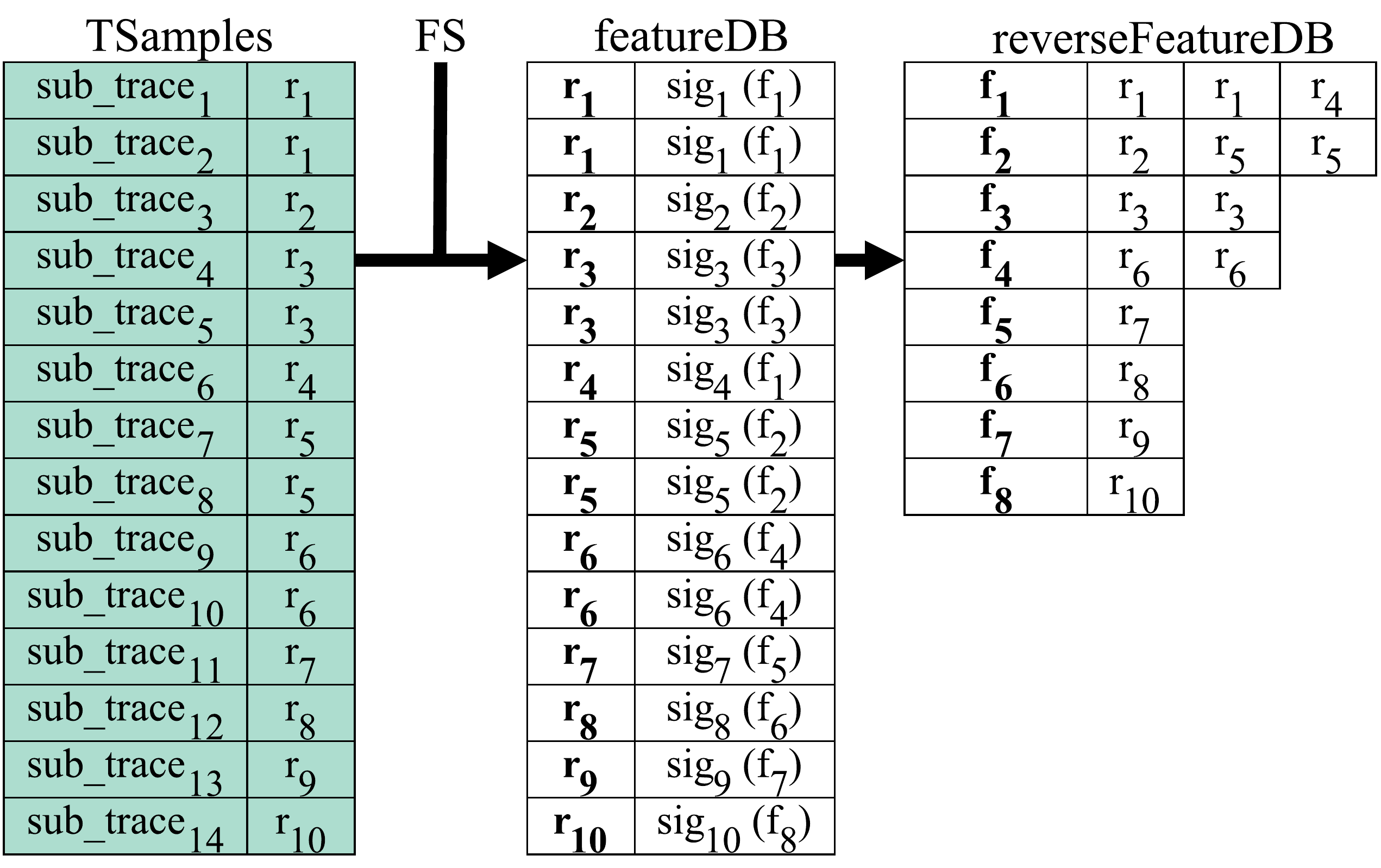}
    \captionsetup{justification=centering}
    \caption{Constructing featureDB and reverseFeatureDB from TSamples and FS}
    \label{fig:featuredb}
    \vspace{-15pt}
\end{figure}

\noindent \textbf{(1) Characterizing tracking cookies.} Tracking cookie is an integral part of the source code of a webpage that records the sequence of webpages previously visited by a user. To account for the impact of tracking cookies on the size of encrypted resource $r_i$, we model the parameter $c_t(r_i)$. 
Note that there is no tracking cookie during profiling time, since each webpage is accessed in an individual session. The possible variation in feature values that arises due to embedded tracking cookies in a user trace is calculated by parsing the source code of the web resource (if it is in text format). Recently developed tools such as CookieCheck~\cite{trevisan20194} can also be used for automated detection of tracking cookies in web resources. Note that only resources that are in text format carry tracking cookies.
For each web resource that carries a tracking cookie, the COMPUTE\_COOKIE\_VAR procedure first identifies the set of URLs from which a user can navigate to the resource. Subsequently, for each of these URLs, it first computes the size of the resource {\em with} the tracking cookie when it is not encrypted. To compute the corresponding resource size when encrypted, we use linear interpolation of known plain-text and encrypted-text pairs. Finally, for each resource $r_i$, the estimated variation is stored as
\begin{equation}
\begin{split}
c_t(r_i)=\{<URL_1,&c_t(r_i,1)>, <URL_2,c_t(r_i,2)>,\\
&\ldots,<URL_x,c_t(r_i,x)>\}.
\end{split}
\end{equation}
Here, $c_t(r_i,j) (1\leqslant j\leqslant x)$ denotes the increase in encrypted size of resource $r_i$ due to an embedded tracking cookie carrying the string $URL_j$.

\noindent\textbf{(2) Characterizing session cookies.} Session cookies are transmitted when a page is accessed for the first time in a session. During profiling, the resources will always have the session cookie. The parameter $c_s(r_i)$ accounts for the discrepancy in the encrypted size of resource $r_i$ due to the possible absence of session cookie in the real user's trace. This happens when the resource is not the first resource to be accessed by the real user. The value of $c_s(r_i)$ can be estimated by considering all the information carried in the header field (e.g., user ID, user agent, max-age of the cookie, its expiry date, etc.) and their possible values. Most of these information are of fixed length, except for the user agent field that contains the browser and OS name. Much like tracking cookies, the COMPUTE\_COOKIE\_VAR stores this information for each resource $r_i$ as
\begin{equation}
\begin{split}
c_s(r_i)=\{<bo_1,&c_s(r_i,1)>,<bo_2,c_s(r_i,2)>,\\
&\ldots,<bo_x,c_s(r_i,x)>\}. 
\end{split}
\end{equation}
 Here $c_s(r_i,j) (1\leqslant j\leqslant x)$ denotes the increase in encrypted size of resource $r_i$ due to the session cookie carrying browser-OS identifier string $bo_j$.

The cookie induced variations thus estimated, are stored in a lookup-table $cookie\_var$, where each entry is of the form $<r_i, \{c_t(r_i), c_s(r_i)\}>$. For resources that do not contain tracking cookies, $c_t(r_i)$ contains $NULL$ value.

%%%%%%%%%%%%%%% Prediction %%%%%%%%%%%%%%%%%%%%%%%%%%%
\subsection{Webpage Prediction Module}\label{sec:analysis}
The \emph{Webpage Prediction Module} of Snoopy takes an encrypted traffic trace $T$ as input, as shown in Step 3 of Figure~\ref{fig:snoopy-components}, and predicts the webpages that are accessed in $T$ using the information stored in the Snoopy database. Snoopy performs this prediction in two steps. First, it predicts the web resources present in $T$. The sequence of predicted resources is denoted as the $predicted\_resources$ sequence. Next, Snoopy uses this sequence to predict the final set of webpages, denoted as $predicted\_webpages$, which it returns to the adversary (Step 4 of Figure~\ref{fig:snoopy-components}).

\subsubsection{Predicting Web Resource Sequence}
For calculating $predicted\_resources$ in $T$, Snoopy first splits the input trace $T$ into sub-traces, each corresponding to an encrypted resource.
From each sub-trace, the feature values are extracted and stored, in the order they appear in $T$, into an array $FValues$. 
The $F$ considered, i.e., encrypted resource size, is obtained using techniques discussed in Section~\ref{sec:profiling}.

As a next step, Snoopy has to identify the actual web resource that may correspond to each feature value in $FValues$, in order of their occurrence in T. It does so by looking-up for each of the $FValues[i]$, its matching resource in the dictionary $reverseFeatureDB$. 
Apart from the fact that multiple resources may have the same feature value, the look-up procedure is also not straightforward due to variations in feature values induced by tracking and session cookies. 
We illustrate the look-up procedure with a concrete example. 

\begin{figure}[t]
    \centering
    \includegraphics[scale=0.2]{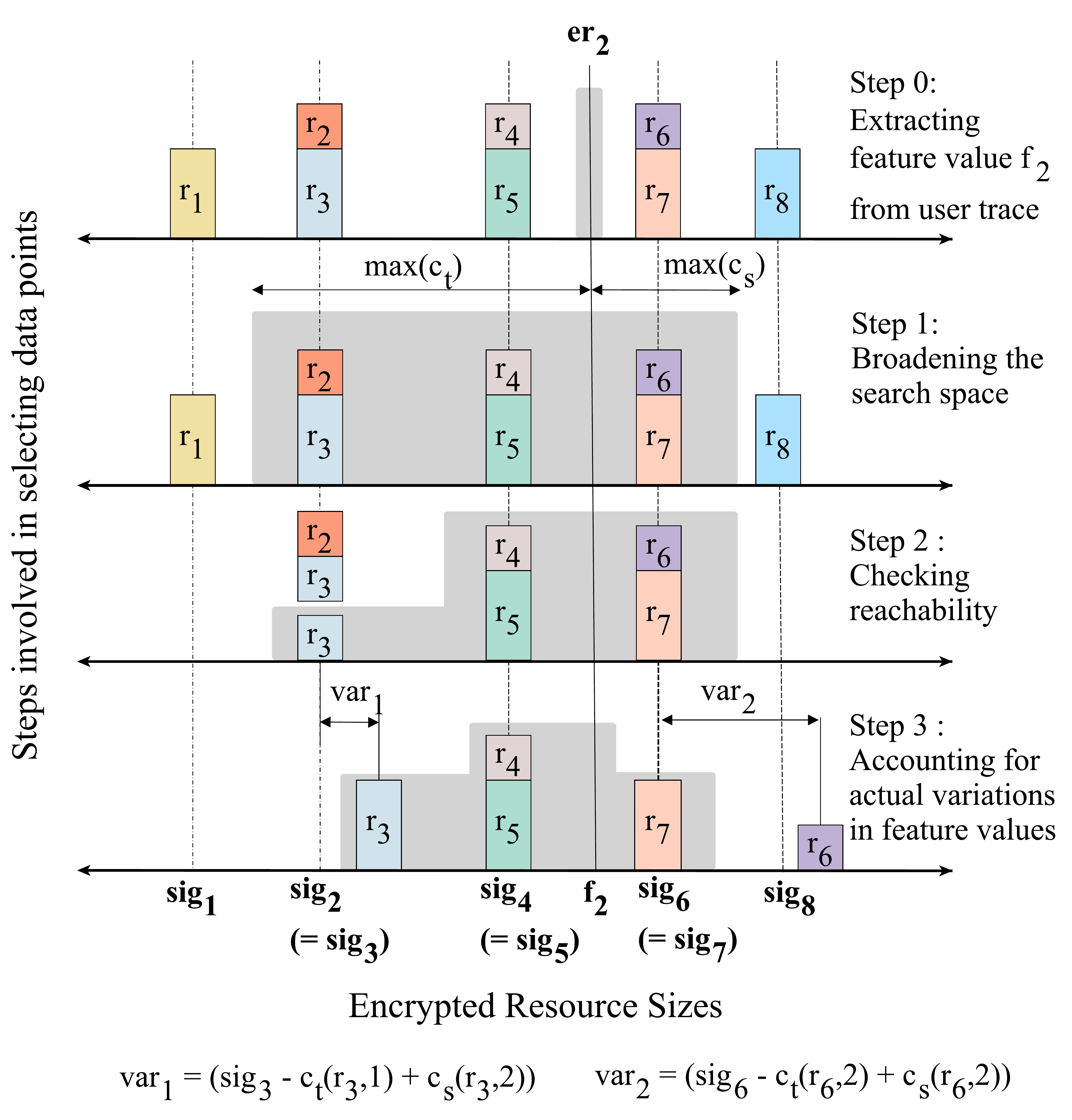}
    \caption{Steps involved in selecting resources that might correspond to a given feature value}
    \label{fig:prediction}
    \vspace{-15pt}
\end{figure}

\noindent {\bf Example:} Consider two encrypted resources $er_1$ and $er_2$ extracted from $T$ such that $er_1$ is accessed by the user before $er_2$. Let $f_1$ and $f_2$ be the $FValues$ of $er_1$ and $er_2$ respectively. The Resource Prediction Module of Snoopy processes $f_1$ followed by $f_2$ in sequential order. We assume without loss of generality that Snoopy has correctly predicted that the resource, viz., $r_4$, corresponds to $er_1$ after processing $f_1$. We now describe how the Resource Prediction Module of Snoopy processes $f_2$. 
Figure~\ref{fig:prediction} shows a stacked bar plot where the X-axis represents feature values (encrypted resource sizes) and the Y-axis represents the frequency of occurrence of different resources in the website (e.g., $r_7$ and $r_6$ have the same size but $r_7$ occurs more than $r_6$).

%%%%%%%%%%%% steps used for the look-up %%%%%%%%%%%%%%%%
\noindent {{\bf Step 1: }{\bf Broadening the search space --}} The value $f_2$ may or may not correspond to the signature of a resource in the database. For instance, Step 0 of Figure~\ref{fig:prediction} shows a scenario where $f_2$ does not match any of the encrypted resource sizes from $reverseFeatureDB$. 
%In fact, the value $f_2$ may be lesser or greater than the value stored in $reverseFeatureDB$. 
This happens due to the differences in the feature value induced by the possible exclusion of session cookies and inclusion of tracking cookies (refer Section~\ref{sec:feature}). %To handle this scenario, the Webpage Prediction Module of Snoopy uses the meta information stored in the $cookie\_var$ database.
If $f_2$ is inclusive of the tracking cookies, it needs to be decremented (by $c_t(r_i,j)$) before performing a $reverseFeatureDB$ look-up. Likewise, if $f_2$ does not contain session cookies, it needs to be incremented (by $c_s(r_i,k)$) before performing a look-up. %Therefore, Snoopy needs to decrement or increment $f_2$ appropriately before performing a lookup. 
However, the presence or absence of both session cookies and tracking cookies for a user cannot be determined by the adversary since it does not have the knowledge of webpages previously accessed by the user in the trace. 
That is, both $c_t(r_i,j)$ and $c_s(r_i,k)$ cannot be determined since $i$ (i.e., resource ID), $j$ (i.e., previously accessed URL), and $k$ (i.e., user agent string) are not known. Instead, we use the maximum possible values for $c_t(r_i,j)$ and $c_s(r_i,k)$, retrieved from $cookie\_var$. Therefore, Snoopy performs a look-up of all values in the range $[(f_2-max(c_t)),(f_2+max(c_s))]$ and adds all the retrieved resources to the \textit{multi}set $relevant\_resources$.

\begin{sloppypar}
Step 1 of Figure~\ref{fig:prediction} shows that although $f_2$ is closest to the sizes of $r_4$, $r_5$, $r_6$, and $r_7$, it could also possibly belong to $r_2$ or $r_3$ due to the aforementioned variations. These resources form the multiset $relevant\_resources=\{r_2, r_3, r_3, r_4, r_5, r_5, r_6, r_7, r_7\}$, and is indicated by the shaded region in Step 1 of Figure~\ref{fig:prediction}.
\end{sloppypar}

\noindent \textbf{Step 2: Reachability checking --} In the real-world, a user starts browsing from a particular webpage and navigates using links embedded in the initial and subsequent webpages. We leverage this realistic behavior to improve the predictions made by Snoopy. In essence, based on the knowledge of previously accessed resources in the trace $T$, the resources in $relevant\_resources$ corresponding to webpages that are not navigable from the previously accessed webpages are eliminated. 
To check the reachability, Snoopy leverages the website structure $\mathbb{G}$ and the resource map $\mathbb{RM}$. 
While performing reachability checking, unlike prior works~\cite{chapman2011automated,miller2014know}, Snoopy considers multi-tab browsing too, i.e., a user can navigate to a webpage that is directly accessible from \textit{any} of the tabs open in the browser at a given point of time. Therefore, Snoopy eliminates only those resources that are not accessible from \textit{any} of the previously accessed webpages in $T$. \hlt{Elimination of unreachable resources further narrows down the search for the actual resource accessed, thereby reducing the computational overheads of Snoopy in the later steps.}

In the example, $er_1$ ($=r_4$) is the previously accessed resource, which is uniquely associated to the webpage $w_3$ (Refer to Figure~\ref{fig:resmap}). Based on this fact, we can infer that the resource $er_2$ can come from either of $w_3$, $w_4$, or $w_5$ and not from $w_1$ or $w_2$. If any resource in the $relevant\_resources$ set that are associated with $w_1$ (we do not have $r_1$ in this example) or $w_2$ (i.e., $r_2$ and $r_3$), its frequency should be reduced accordingly. 
Therefore, the set of reachable resources now contains $\{r_3,r_4,r_5,r_5,r_6,r_7,r_7\}$, denoted by $reachable\_resources$, indicated by the shaded region in Step 2 of Figure~\ref{fig:prediction}, which is scaled-down as compared to that in Step 1. 

\noindent\textbf{Step 3: Accounting for cookie-induced variations in reachable resources --} \hlt{Next, Snoopy computes $(sig_i - c_t(r_i,x) + c_s(r_i,y))$ to estimate the actual encrypted feature value for each resource $r_i$ in $reachable\_resources$, with the knowledge that the previously accessed resource was $r_4$. For this, Snoopy estimates \textbf{(1)} the URL contained in its tracking cookie, if $c_t(r_i)$ is not $NULL$. This URL (say, $URL_x$) can be estimated using the technique used for reachability checking; and, \textbf{(2)} the browser and OS information contained in the session cookie in the Application layer header. This information (say, $bo_y$), can be estimated by using existing browser fingerprinting techniques~\cite{husak2016https} and OS fingerprinting techniques~\cite{lavstovivcka2020using}. From this knowledge of $URL_x$ and $bo_y$, Snoopy retrieves the corresponding $c_t(r_i,x)$ and $c_s(r_i,y)$ respectively from the $cookie\_var$ table.

This step further narrows down the search space.
This process is illustrated in Step 3 of Figure~\ref{fig:prediction}, which shows the case where the resource $r_6$ is eliminated from the set $reachable\_resources$. In this example, $r_4$, $r_5$ and $r_7$ do not have cookie induced variation in feature values.
}

\noindent\textbf{Step 4: Assigning Weightage to Resources --} 
In this step, Snoopy identifies the resource that is most likely to correspond to $er_2$. For this, Snoopy assigns a weightage to each resource in $reachable\_resources$ based on the number of instances of each resource in the multiset and the differences between their updated feature values and $f_2$. For each unique resource $r_i\in reachable\_resources$, we denote the distance between its updated feature value (from Step 3) and $f_i$ as $diff_i$ and the frequency of $r_i$ in $reachable\_resources$ as $freq_i$. The weightage $w_i$ assigned by Snoopy to resource $r_i$ is $\dfrac{freq_i}{\mid diff_i\mid}$, where $diff_i \neq 0$. 
Finally, Snoopy considers the resource with the highest weightage to be the one corresponding to $er_2$.

\subsubsection{Predicting Webpages Accessed by the User}
Snoopy uses a sequence extraction algorithm for predicting accessed webpages from the $predicted\_resources$ sequence. Snoopy starts with the first resource (say, $r_{p1}$) in $predicted\_resources$, and uses $\mathbb{RM}$ to identify all the webpages that contain $r_{p1}$. Next, using the sequence set $\mathbb{S}$, Snoopy identifies all the webpages where $r_{p1}$ is the first resource. For each of these webpages, Snoopy then checks whose resource download sequence forms a subsequence of the $predicted\_resources$ sequence. When Snoopy identifies the webpage, it adds the corresponding webpage in the $predicted\_webpages$ set. In case Snoopy is able to extract multiple resource download sequences from $predicted\_resources$, it selects the one whose elements appear closer to each other as compared to the others. Once Snoopy extracts a sequence from $predicted\_resources$, it removes them from $predicted\_resources$ and continues the above process starting from the new first resource in the sequence. This process continues until the $predicted\_resources$ sequence is empty. \hltr{In the latter rounds of sequence extraction, Snoopy eliminates those resources that are cache-able \textit{and} have already been extracted from consideration.} This handles the scenario where certain resources had not been downloaded multiple times due to web caching being enabled. Note that, more sophisticated sequence extraction algorithms could be used in Snoopy in future. However, our seemingly naive algorithm is not a bottleneck at the moment. As we show in Section~\ref{sec:evaluation}, using this algorithm, Snoopy achieved a considerably high prediction accuracy on the websites that we used for our experiments.

%%%%%%%%%%%%%%%%%%%%%%%%%%%%%%%%%%%%%%%%%%%%%%%%%%%%%%%%%%%%%%%%%%%%%%%%%%%%%%%%%%%%%%%%%%%%%%%%%%%%%%%%%%%%%%%%%%%%%%%%%%%%%%%%%%%%

\section{Implementation and Evaluation}\label{sec:evaluation}
We first describe our implementation and define the metrics used for evaluating Snoopy. Thereafter, we describe the experimental setup, which includes the websites used for evaluation, the user browsing scenarios considered for evaluation, and the traffic sample collection methodology. Subsequently, we evaluate Snoopy by answering a set of research questions. 
%In this section, we perform a comprehensive evaluation of Snoopy. We not only evaluate Snoopy on the basis of its prediction accuracy, but we also evaluate its generalization capabilities across each factor (user interest, user behavior, user preference, network conditions) individually, in order to demonstrate its suitability for mass-scale surveillance. Additionally, we also evaluate the extent to which Snoopy complies with a finite query model. First, we provide a detailed description of our experimental setup. Next, we show the prediction accuracy of Snoopy when users of real websites access the webpages in diverse browsing contexts. Finally, we present the generalization capabilities of Snoopy.
\subsection{Implementation}\label{sec:implementation}
We implemented the functional modules of Snoopy as Python libraries. 
\hlt{For comparison with existing works~\cite{chapman2011automated,cai2012touching,yan2018feature,miller2014know,liberatore2006inferring,hayes2016k,panchenko2016website}, we implemented some of the state-of-the-art works on webpage fingerprinting to the best of our abilities with the help of open-source Python libraries~\cite{existing-code}.} Encrypted network traffic traces used for webpage fingerprinting and testing was captured using TShark(v2.6.6) packet capture tool, which was running on the network gateway, although the same could be done by the attacker at any intermediate network router in the real world. 

\subsection{Metrics}\label{sec:metrics}
% The goal of webpage identification techniques in the context of mass-scale surveillance is to achieve a high prediction accuracy across multiple browsing scenarios, while adhering to a finite query model for practicality. 
We define the metrics that we use for evaluating Snoopy and existing state-of-the-art webpage identification techniques as follows:
\begin{description}
    \item [Generalization factors ($\mathbf{GF}$):] The set of factors related to user's browsing behavior that the webpage identification technique aims to generalize across. $\mathbf{GF} \subseteq \{I,BC,T,O,B,N\}$, where $I$ denotes user interest and thereby the set of webpages in the website that the attacker needs to fingerprint, $BC$ denotes the set of browser configurations (cache and cookie settings) considered by the attacker, $T$ denotes the number of tabs used by a user in a browsing session, $O$ indicates the set of Operating Systems that the attacker generalizes across, $B$ denotes the set of Browsers, and $N$ denotes network conditions.
    \item[{Number of Queries} ($\mathbf{N_q}$):] The maximum number of website queries (accesses) that an attacker can perform on the targeted website for collecting traffic samples for training (webpage profiling).
    % allowed required by a webpage identification technique to fingerprint the webpages of a target website.
     \item [Fingerprinting Accuracy ($\mathbf{FA}$):] The percentage of webpages accessed by a user (of the target website) in a browsing session that are correctly identified by a webpage identification technique.  
\end{description}

\begin{table}[t]
\centering
{\small
% \resizebox{\textwidth}{!}
{%
\begin{tabular}{|l|l|l|l|l|l|}
\hline
\rowcolor[HTML]{EFEFEF} 
\multicolumn{1}{|c|}{\cellcolor[HTML]{EFEFEF}\scriptsize{\begin{tabular}[c]{@{}c@{}}Sl \\ No\end{tabular}}} & \multicolumn{1}{c|}{\cellcolor[HTML]{EFEFEF}\scriptsize{Website}} & \multicolumn{1}{c|}{\cellcolor[HTML]{EFEFEF}\scriptsize{\begin{tabular}[c]{@{}c@{}}HTTP \\ version\end{tabular}}} & \multicolumn{1}{c|}{\cellcolor[HTML]{EFEFEF}\scriptsize{\begin{tabular}[c]{@{}c@{}}TLS \\ version\end{tabular}}} & \multicolumn{1}{c|}{\cellcolor[HTML]{EFEFEF}\scriptsize{\begin{tabular}[c]{@{}c@{}}No of \\ pages\end{tabular}}} & \multicolumn{1}{c|}{\cellcolor[HTML]{EFEFEF}\scriptsize{\begin{tabular}[c]{@{}c@{}}Type of \\ webpages\end{tabular}}} \\ \hline
1 & RS & 1.1 & 1.2 & 20 & plain HTML \\ \hline
2 & SBC & 2.0 & 1.3 & 27 & HTML scripts \\ \hline
3 & IC\_1 & 1.1 & 1.2 & 95 & aspx \\ \hline
% \rowcolor[HTML]{FFFFC7} 
4 & B\_1 & 1.1 & 1.2 & 444 & aspx \\ \hline
% \rowcolor[HTML]{FFFFC7} 
5 & B\_2 & 1.1 & 1.2 & 458 & HTML scripts \\ \hline
% \rowcolor[HTML]{FFFFC7} 
6 & B\_3 & 1.1 & 1.2 & 549 & aspx \\ \hline
% \rowcolor[HTML]{DAF7D4}
7 & B\_4 & 1.1 & 1.2 & 965 & HTML + JS \\ \hline
% \rowcolor[HTML]{DAF7D4} 
8 & IC\_2 & 1.1 & 1.2 & 849 & HTML + JS \\ \hline
% \rowcolor[HTML]{DAF7D4} 
9 & B\_5 & 1.1 & 1.2 & 1964 & plain HTML \\ \hline
10 & PS & 2.0 & 1.2 & 40,323 & HTML + JS \\ \hline
\end{tabular}%
}
\caption{Characteristics of Profiled Websites\\ {\scriptsize RS - Retail Store website, SBC - Service Based Company website, IC - Insurance Company website, B - Bank website, PS - Political Survey website}}
\label{websitech}
}
\vspace{-15pt}
\end{table}

\subsection{Experimental setup}\label{sec:expsetup}
\noindent \textbf{Websites Used.} 
We evaluate Snoopy and related works~\cite{liberatore2006inferring,chapman2011automated,cai2012touching,miller2014know,hayes2016k,panchenko2016website,yan2018feature} on $20$ websites that include some of the Fortune $100$ companies, financial organizations, and service-based companies from different parts of the world. The scale of our experiments is at par with existing works on webpage identification~\cite{cheng1998traffic,sun2002statistical,danezis2010traffic,miller2014know,hayes2016k,zhuo2017website,xu2018multi,shen2019webpage,alan2019client}. As mentioned earlier, this paper deals with \textit{webpage identification} performed {\em on a website} of interest to the adversary, in contrast to website identification that concerns thousands of websites.
Due to space constraints, we present the results for ten of these websites, listed in Table~\ref{websitech}, that %encompass a wide range of web technologies, sizes, and business domains and 
are representative of the entire set. 
The names of these websites are anonymized to protect the websites from being targets of this attack. 

\noindent \textbf{User Browsing Scenarios.} %We evaluate Snoopy and existing webpage identification techniques on a diverse set of browsing scenarios to account for mass-scale surveillance. 
We consider different combinations of \textbf{(1)} Operating Systems viz.; Ubuntu $18.04$, and  Microsoft Windows $7$, \textbf{(2)} Browsers viz.; Mozilla Firefox $63.0.3$, and Google Chrome $67.0$, \textbf{(3)} cache settings viz.; ON and OFF, \textbf{(4)} cookie preferences viz.; Allowed and Prohibited, and \textbf{(5)} network conditions. Users can start browsing from any webpage in the website that they would be interested in, and browse the pages in any order. Further, we also allow users to freely browse up to $15$ different webpages in each browsing session, either sequentially or by using multiple parallel tabs, with no restriction on the transition time from one webpage to another. Most existing works do not consider cases where the users browse more than one page in a session and even if they do, they restrict it to a small number(for e.g., $4$ in ~\cite{xu2018multi} and $2$ in ~\cite{gu2015novel}). \hlt{These numbers are much lower than the average number of parallel browser tabs used by website visitors, as pointed out in a study\footnote{Mozilla has removed the dataset compiled for the Test Pilot study conducted in 2010 from public domains. But we can speculate that the trend of tabbed webpage browsing has only gone up in the last decade.} conducted by Mozilla~\cite{mozillatabs}.}

\noindent \textbf{Traffic Sample Collection.} 
We developed a Python bot for collecting encrypted traffic samples for evaluating Snoopy as well as the existing works. Our traffic collection bot used Selenium for simulating behavior of real website users while collecting test traffic samples. For creating webpage fingerprints, our bot accessed the webpages as many times as required by the different webpage fingerprinting techniques, within the limits of the finite query model. \hlt{Depending on the requirements of our experiments, the bot performs either sequential or single-page traffic sample collection.} We introduced a delay of $1$ minute between subsequent browsing sessions. We observed that repeated website accesses from the same network to the same website needed to be separated by this time in order for it to not be flagged and blocked by the website. 
% Recall that, based on the characteristics of training and testing traffic samples, webpage identification techniques can be classified into $3$ types - \textbf{(1)} Techniques that use single webpage traffic for training and single webpage traffic for testing~\cite{}, \textbf{(2)} Techniques that use sequential webpage access traffic for training and sequential webpage access traffic for testing~\cite{}, and \textbf{(3)} Techniques that use single webpage traffic for training and sequential webpage access traffic for testing~\cite{}. 
\hlt{For evaluations that required sequential browsing traffic samples, the browsing sequence length was varied from $3$ to $15$ webpages per session. For evaluations that required traffic samples collected over various network conditions, the traffic sample collection was conducted over several months and from different geographical locations to ensure variations in network conditions. In addition to the webpage(s) accessed in a browsing session (which is needed for computing the accuracy), our bot also recorded the sequence of web objects downloaded in each browsing session, which we use for additional evaluation.}

\subsection{Results}\label{results}
In this section, we first evaluate the suitability of Snoopy in the context of practical mass-surveillance. In the prior sections, we have discussed about the two key requirements for conducting practical mass-surveillance -- generalization and compliance with a finite query model. Therefore, we primarily evaluate Snoopy on these two parameters, through experiments that intend to answer the following questions.
\begin{itemize}[leftmargin=.4in,rightmargin=.3in]
    % \item [\textbf{(Q1)}] To what extent can Snoopy generalize in practical browsing scenarios? 
    \item [\textbf{(Q1)}] How effective is Snoopy in different browsing contexts? (Refer to Section~\ref{sec:diverse-browsing})
    % predicting the browsing activities of users  average prediction accuracy of Snoopy across different browsing contexts?
    \item [\textbf{(Q2)}] How is the effectiveness of Snoopy and related works affected with changes in the value of $N_q$ (number of queries allowed)? (Refer to Section~\ref{sec:fqm})
    % To what extent does Snoopy comply with a finite query model?
    \item [\textbf{(Q3)}] How well does Snoopy generalize across user interests? (Refer to Section~\ref{sec:user-interests})
    \item [\textbf{(Q4)}] How well does Snoopy generalize across different configurations of a browser? (Refer to Section~\ref{sec:browser-config})
    \item [\textbf{(Q5)}] How well does Snoopy generalize in terms of number of tabs used by the end user while browsing? (Refer to Section~\ref{sec:multi-tab})
    \item [\textbf{(Q6)}] How well does Snoopy generalize in terms of variations in Operating Systems? (Refer to Section~\ref{sec:os})
    \item [\textbf{(Q7)}] How well does Snoopy generalize in terms of variations in Browsers? (Refer to Section~\ref{sec:browser})
    \item [\textbf{(Q8)}] How well does Snoopy generalize across various network conditions? (Refer to Section~\ref{sec:network})
\end{itemize}

\subsubsection{Average prediction accuracy of Snoopy across diverse browsing scenarios}\label{sec:diverse-browsing}
We now show the prediction accuracy ($FA$) of Snoopy when it builds webpage fingerprints (trains) for any one browsing scenario and uses this information to predict user activities across diverse browsing scenarios. For this experiment, we considered all the $10$ websites. \hlt{For each website, we built a training dataset where we consider $GF=\{I,BC,T,O,B,N\}$ such that, $I$=\{all webpages in the website\}, $BC=\{\{$Caching ON, Cookies Allowed$\},\{$,Caching OFF, Cookies Prohibited$\}\}$, $T=1$, $O=\{Ubuntu\}$, $B=\{Firefox\}$. $N$ was kept constant by collecting all traffic samples from the same system within a short span of time. We collect $10$ such traffic samples for each webpage. 

For each website, we also build a test dataset, where we consider $I$=\{all webpages in the website\}, vary $BC$ across the values $\{\{$Caching ON, Cookies Allowed$\},\{$Caching ON, Cookies Prohibited$\},\{$Caching OFF, Cookies Allowed$\},\{$,Caching OFF, Cookies Prohibited$\}\}$, vary $T$ from $1$ to $15$, vary $O$ across the values $\{Windows, Ubuntu\}$, and vary $B$ across the values $\{Chrome, Firefox\}$. The network conditions $N$ were varied by collecting traffic samples over several months from different geographical locations (refer to the test dataset described in Section~\ref{sec:expsetup}).} 

% For each website, the training dataset consists of $10$ traffic samples from each webpage that were collected using Ubuntu OS and Firefox Browser, with cookies Allowed, and caching set to ON and OFF alternately. The webpages were opened on one browser tab at a time. The test data comprised $200$ instances of encrypted browsing traffic from each website. There were equal number of traffic samples from all possible combinations of elements in $GF=\{I,BC,T,O,B,N\}$, where $I$ was the set of all webpages of the website of interest, $BC=\{\{$Caching ON, Cookies Allowed$\},\{$Caching ON, Cookies Prohibited$\},\{$Caching OFF, Cookies Allowed$\},\{$,Caching OFF, Cookies Prohibited$\}\}$, $T$ was varied from $1$ to $15$, $O=\{Windows, Ubuntu\}$, and $B=\{Chrome, Firefox\}$. The network conditions were inherently varied since the traffic samples were collected over several months from different geographical locations (refer to the test dataset described in Section~\ref{sec:expsetup}). 

Table~\ref{table:predaccuracy1} shows the webpage prediction accuracy ($FA$) of Snoopy on the $10$ websites. The $FA$ value is more than $90\%$ for most of the websites, indicating the ability of Snoopy to generalize across different browsing contexts. \hlt{In the rest of this section, we show how the different factors in $GF$ individually influence the $FA$ value of Snoopy, when subjected to constraints on the value of $N_q$, the maximum number of queries allowed to a website.} 
\begin{table}[t]
\renewcommand{\tabcolsep}{2pt}
\centering
\begin{tabular}{|c|c|c|c|c|c|c|}
\hline
\cellcolor[HTML]{EFEFEF} & \cellcolor[HTML]{EFEFEF} &  \multicolumn{3}{c|}{\cellcolor[HTML]{EFEFEF}\scriptsize{\begin{tabular}[c]{@{}c@{}}Webpage identification accuracy\end{tabular}}} \\ \cline{3-5} 
\cellcolor[HTML]{EFEFEF} & \cellcolor[HTML]{EFEFEF} & \cellcolor[HTML]{EFEFEF} & \cellcolor[HTML]{EFEFEF} & \cellcolor[HTML]{EFEFEF} \\
\multirow{-3}{*}{\cellcolor[HTML]{EFEFEF}\scriptsize{Website}} & \multirow{-3}{*}{\cellcolor[HTML]{EFEFEF}\scriptsize{\begin{tabular}[c]{@{}c@{}}No. of\\ webpages\end{tabular}}} &  \multirow{-2}{*}{\cellcolor[HTML]{EFEFEF}\scriptsize{\begin{tabular}[c]{@{}c@{}}Accurately\\ identified (\%)\end{tabular}}} & \multirow{-2}{*}{\cellcolor[HTML]{EFEFEF}\scriptsize{\begin{tabular}[c]{@{}c@{}}Not\\ identified (\%)\end{tabular}}} & \multirow{-2}{*}{\cellcolor[HTML]{EFEFEF}\scriptsize{\begin{tabular}[c]{@{}c@{}}Wrongly\\ identified (\%)\end{tabular}}}\\ \hline
\begin{tabular}[c]{@{}c@{}}IC\_1\end{tabular} & 95 & 93 & 7 & 0\\ \hline
IC\_2 & 849 & 89 & 11 & 0 \\ \hline
B\_1 & 444 & 99 & 1 & 0\\ \hline
B\_2 & 458 & 99 & 1 & 0\\ \hline
B\_3 & 549 & 97 & 0 & 3\\ \hline
B\_4 & 965 & 88 & 11 & 1\\ \hline
B\_5 & 1964 & 90 & 3 & 7\\ \hline
SBC & 27 & 81 & 7 & 12\\ \hline
RS & 20 & 75 & 21 & 4\\ \hline
PS & 40323 & 83 & 17 & 0\\ \hline
\end{tabular}%
\caption{Webpage prediction accuracy of Snoopy}
\label{table:predaccuracy1}
% \vspace{-15pt}
\end{table}

\subsubsection{Compliance with a finite query model: Snoopy vs. ML-based solutions}\label{sec:fqm}

\hlt{We now evaluate Snoopy on its ability to comply with a finite query model with respect to existing works. For this experiment, we considered the website $B\_4$ and generalization factors $GF=\{I,BC,T,O,B,N\}$, where $I$ is a set of $200$ random webpages of the website $B\_4$, $BC=\{\{$Caching OFF, Cookies Allowed$\}\}$, $T=1$, $O=\{Ubuntu\}$, $B=\{Firefox\}$. $N$ was constant since the traffic samples were collected within a short span of time. Our test dataset comprised traffic traces from the website $B\_4$ collected using the above configuration. We built three different training datasets with traffic samples collected using the same configuration, but varying the number of samples collected per webpage. We set $N_q$, the maximum number of queries allowed, as $w \times s$, where $w$ is the number of webpages to be fingerprinted and $s$ is the number of traffic samples collected per webpage. Keeping $w=200$ constant, we vary the value of $s$ as $s = 10, 5,$ and $3$.} 

\begin{figure}[t]
    \centering
    \includegraphics[scale=0.4]{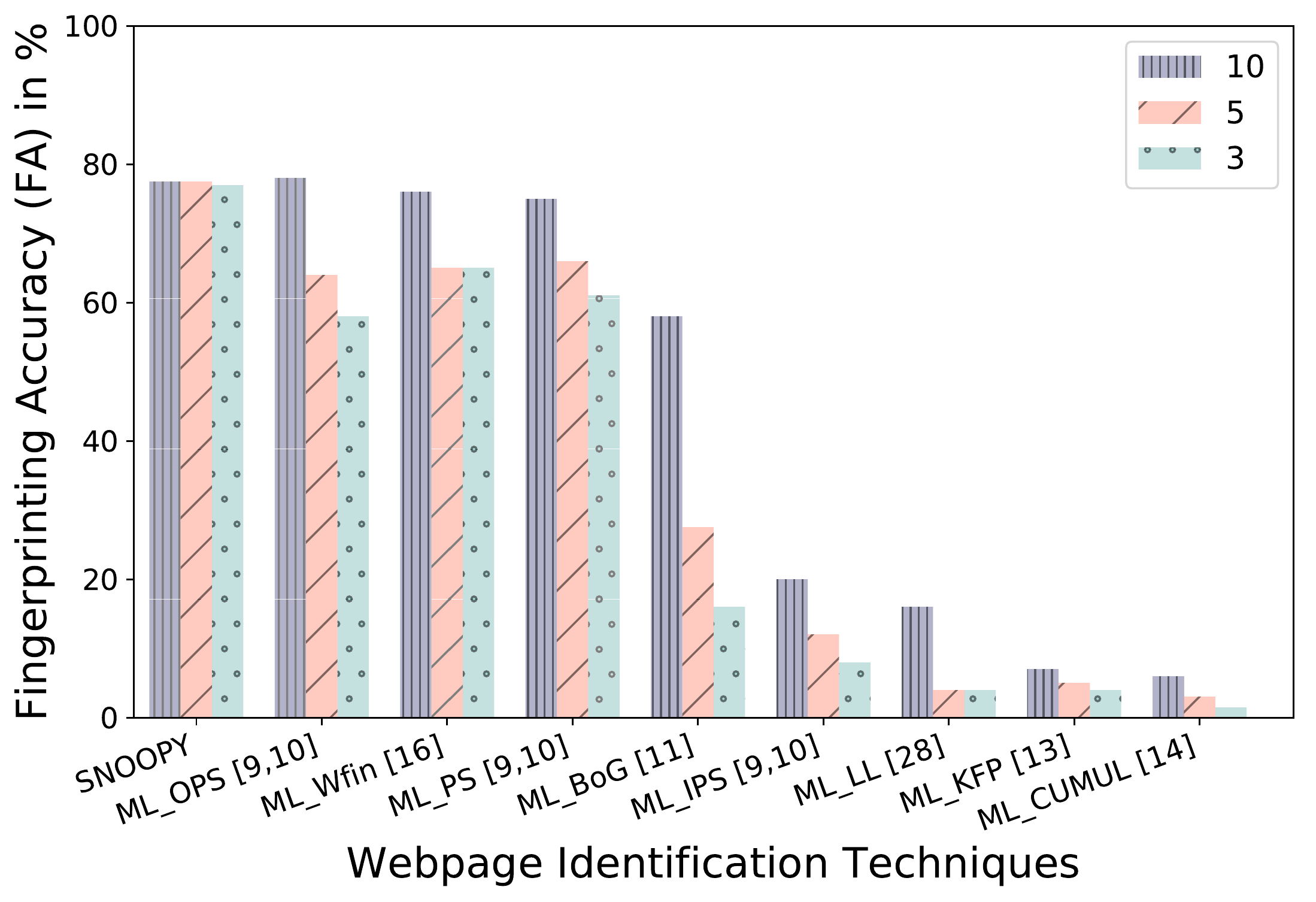}
    \caption{Importance of training set size $N_q$ on fingerprinting accuracy $FA$ (in \%) of existing ML models vs. Snoopy (for 200 web pages)}
    \label{fig:nq-vs-fa}
    % \vspace{-15pt}
\end{figure}

Figure~\ref{fig:nq-vs-fa} plots the fingerprinting accuracy ($FA$) for Snoopy and existing works for different values of $s$. From the figure, we can observe that \textbf{(1)} When $s=10$, the prediction accuracy of Snoopy ($77.75\%$) is at par with the existing ML-based techniques, and; \textbf{(2)} When $s=5$ or $s=3$, we observed a drop in $FA$ of ML techniques (for e.g., from 78\% to 58\% in case of ML\_OPS~\cite{cai2012touching,chapman2011automated}) . On the other hand, we can see that $FA$ for Snoopy remains unchanged even after lowering the value of $s$. \hlt{Since we have seen that for the three different $s$ values considered in this experiment, the ML models have the highest $FA$ value when $s=10$, we use $s=10$ to constrain the value of $N_q$ for the finite query model assumed in the rest of our experiments (shown in Section~\ref{sec:user-interests}--Section~\ref{sec:network}).}
% As we see in the next subsection, the requirement for a high $N_q$ for existing ML-based techniques gets exacerbated with an increase in the number of webpages of interest ($w$).

\subsubsection{Generalization across User Interests}\label{sec:user-interests}
\hlt{We now show how the fingerprinting accuracy ($FA$) of Snoopy and existing works change with the number of webpages of interest ($I$). For this experiment, we consider the website $B\_4$ and generalization factor $GF=\{I,BC,T,O,B,N\}$, where we set $BC=\{\{$Caching OFF, Cookies Allowed$\}\}$, $T=1$, $O=\{Ubuntu\}$, $B=\{Firefox\}$. $N$ was constant since the traffic samples were collected within a short span of time. We varied $I$ as sets of $50, 100,$ and $200$ webpages for building three different training datasets. For this experiment, $N_q=|I|\times10$, which is the constraint imposed by our finite query model. For each of the training datasets, we built test datasets using encrypted traffic traces with the same constraints as the training dataset.}

\begin{figure}[t]
    \centering
    \includegraphics[scale=0.4]{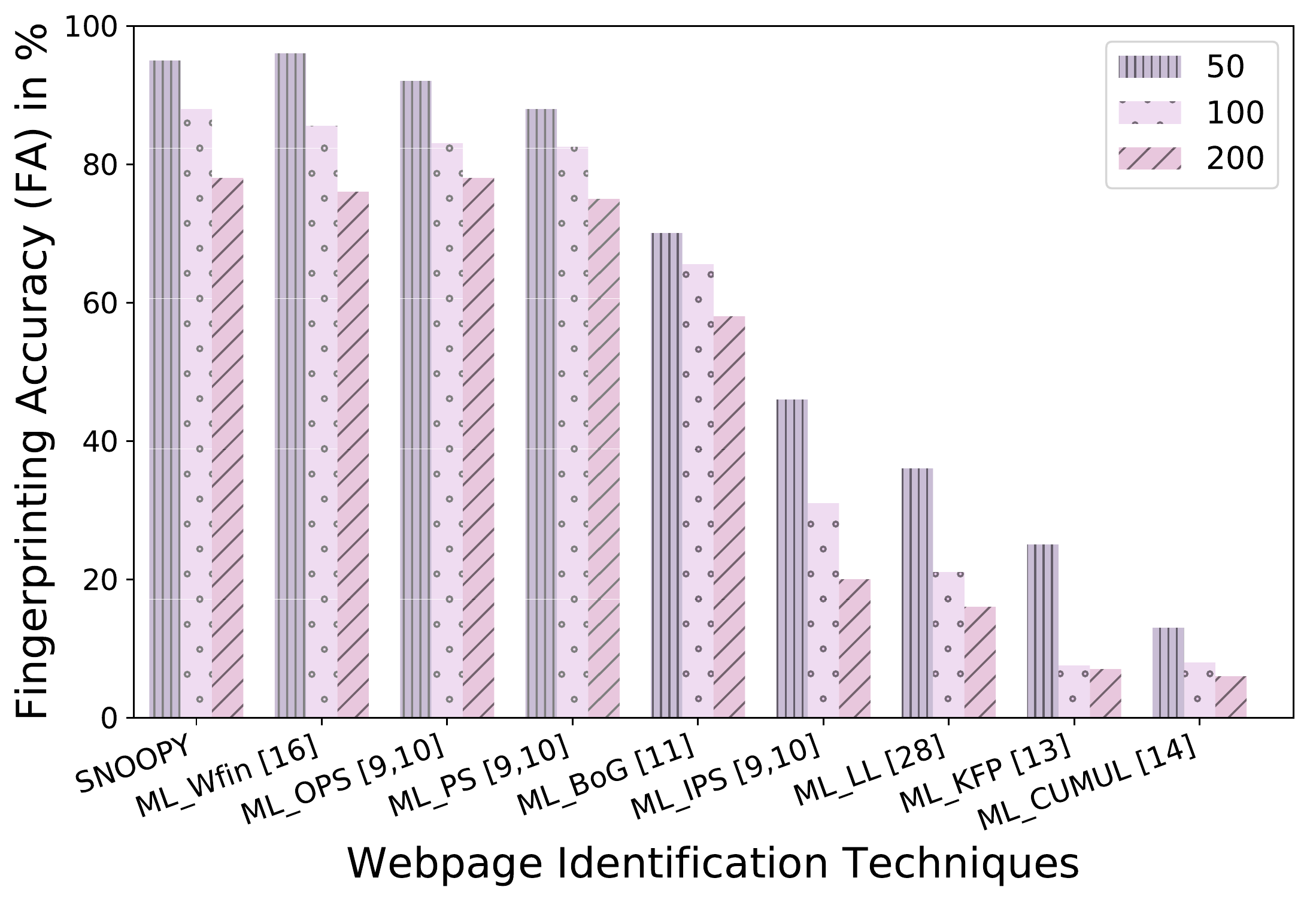}
    \caption{Fingerprinting accuracy $FA$ (in \%) of Snoopy vs. existing techniques for different number of webpages of interest}
    \label{fig:userinterest}
\end{figure}

Figure~\ref{fig:userinterest} shows the fingerprinting accuracy ($FA$) of Snoopy and existing works when the number of webpages of interest increases. From the figure we can observe that \textbf{(1)} Despite using a simple feature and a simple approach, Snoopy gives the same prediction accuracy as some of the best-performing ML-based techniques (e.g., ML\_Wfin~\cite{yan2018feature}), and; \textbf{(2)} When $|I|$ was increased from $50$ to $100$ and $200$, there was a fall in the value of $FA$ for Snoopy as well as the existing ML-based techniques. Even then, Snoopy gave the same prediction accuracy as the best-performing ML-based techniques (for e.g., ML\_Wfin~\cite{yan2018feature} and ML\_OPS~\cite{chapman2011automated,cai2012touching}). 

\hlt{A detailed inspection revealed that the drop in the $FA$ value of Snoopy when the size of $I$ was gradually increased was due to the fact that many of the webpages in the larger sets had similar fingerprints. 
In Section~\ref{sec:resvspage}, we will discuss the reason behind this in more details, and in Section~\ref{sec:snoopy-meets-ml} we will discuss possible ways to improve the fingerprinting accuracy ($FA$) of Snoopy.}  

\subsubsection{Generalization across Browser Configurations}\label{sec:browser-config}
Next, we evaluate Snoopy and existing works on their ability to generalize across different configurations of caching and cookie settings for a given browser ($BC$). For this experiment, we consider the website $B\_4$ and $GF=\{I,BC,T,O,B,N\}$, such that $I$ is a set of $200$ random webpages of the website $B\_4$, $T=1$, $O=\{Ubuntu\}$, $B=\{Firefox\}$, and $N$ was constant.
\hlt{While building our training dataset, we keep $BC=\{$Caching OFF, Cookies Allowed$\}$ constant, and collect $10$ samples of browsing traffic from each of the $200$ webpages of $B\_4$ that we considered. For this experiment, $N_q=200\times 10$, which is the constraint imposed by the finite query model. In our test dataset, we vary $BC$ across the values $\{\{$Caching OFF, Cookies Allowed$\},\{$Caching ON, Cookies Allowed$\}\}$, with all other factors similar to the training dataset.}

% We varied $BC$ for our test dataset across the values $\{\{$Caching OFF, Cookies Allowed$\},\{$Caching ON, Cookies Allowed$\}\}$. However, we built a single training dataset where we set $BC$ to $\{$Caching OFF, Cookies Allowed$\}$. For building the training dataset, we used $10$ samples of browsing traffic from each of the $200$ webpages of $B\_4$ that we considered. For this experiment, $N_q=200\times 10$, which is the constraint imposed by the finite query model.

Table~\ref{tab:browser-config} shows a comparison of the fingerprinting accuracy ($FA$) of Snoopy and related works. For the straightforward case, where training and testing used the same $BC$ configuration, Snoopy achieves an $FA$ that is comparable ($FA\approx 75\%-78\%$) to the best performing ML models as expected. However, when the $BC$ configurations for testing and training were different, Snoopy outperforms even the best ML techniques. For example, Snoopy achieves an $FA=72\%$ as compared to ML\_PS~\cite{chapman2011automated, cai2012touching} that achieves $FA=61\%$.

\begin{table}[t]
\centering
\begin{tabular}{|c|c|c|}
\hline
\rowcolor[HTML]{EFEFEF} 
\cellcolor[HTML]{EFEFEF}{\color[HTML]{000000} } & \multicolumn{2}{c|}{\cellcolor[HTML]{EFEFEF}{\color[HTML]{000000} Browser Cache Configuration}} \\ \cline{2-3} 
\rowcolor[HTML]{EFEFEF} 
\multirow{-2}{*}{\cellcolor[HTML]{EFEFEF}{\color[HTML]{000000} Classifier}} & {\color[HTML]{000000} \begin{tabular}[c]{@{}c@{}}Training: OFF\\ Testing: ON\end{tabular}} & {\color[HTML]{000000} \begin{tabular}[c]{@{}c@{}}Training: OFF\\ Testing: OFF\end{tabular}} \\ \hline
\textbf{SNOOPY} & \textbf{72} & \textbf{78} \\ \hline
ML\_PS~\cite{cai2012touching,chapman2011automated} & 61 & 75 \\ \hline
ML\_OPS~\cite{cai2012touching,chapman2011automated} & 60 & 78 \\ \hline
ML\_Wfin~\cite{yan2018feature} & 51 & 76 \\ \hline
ML\_BoG~\cite{miller2014know} & 50 & 58 \\ \hline
ML\_LL~\cite{liberatore2006inferring} & 14 & 16 \\ \hline
ML\_KFP~\cite{hayes2016k} & 4.5 & 7 \\ \hline
ML\_IPS~\cite{cai2012touching,chapman2011automated} & 3 & 20 \\ \hline
ML\_CUMUL~\cite{panchenko2016website} & 2.5 & 6 \\ \hline
\end{tabular}
% \vspace{5pt}
\caption{Accuracy (in \%) of Snoopy vs. ML models when tested on data points from a different browser configuration (for 200 webpages).} 
\label{tab:browser-config}
\end{table}

\subsubsection{Support for prediction on multi-tab browsing traffic}\label{sec:multi-tab}

\hlt{We now evaluate Snoopy in the context of a real-world scenario where the users of the targeted website open the webpages simultaneously in multiple parallel browser tabs ($T$). For this experiment, we considered the website $B\_4$ and $GF=\{I,BC,T,O,B,N\}$, such that $I$ is a set of $200$ random webpages of the website $B\_4$, $BC=\{$Caching OFF, Cookies Allowed$\}$, $O=\{Ubuntu\}$, $B=\{Firefox\}$, and $N$ was constant. While building our training dataset, we keep $T=1$ constant, and collect $10$ samples of browsing traffic from each of the $200$ webpages of $B\_4$ that we considered. For this experiment, $N_q=200\times 10$, which is the constraint imposed by the finite query model. In our test dataset, we vary $T$ from $3$ to $7$ with all other factors similar to the training dataset.
% For this experiment, we considered the website $B\_4$. For building our test dataset we consider $GF=\{I,BC,T,O,B,N\}$, where we only vary $T$ from $3$ to $7$ with all other factors constant across all the data points. 
% $I$ is the set of all $965$ webpages of $B\_4$, $BC$ = $\{\{$Caching ON, Cookies Allowed$\}\}$, $O=\{Ubuntu\}$, $B=\{Firefox\}$, and $N$ was constant. The test data comprised $200$ encrypted traffic traces corresponding to multi-tab browsing sessions involving random sets of webpages opened in parallel on $3-7$ tabs. We built the training dataset using $10$ single-tab browsing traffic using $Firefox$ browser from each of the $965$ webpages, with the browsing configuration set to $\{\{$Caching ON, Cookies Allowed$\}$, $\{$Caching OFF, Cookies Prohibited$\}\}$.
}

\begin{figure}[t]
    \centering
    \includegraphics[scale=0.5]{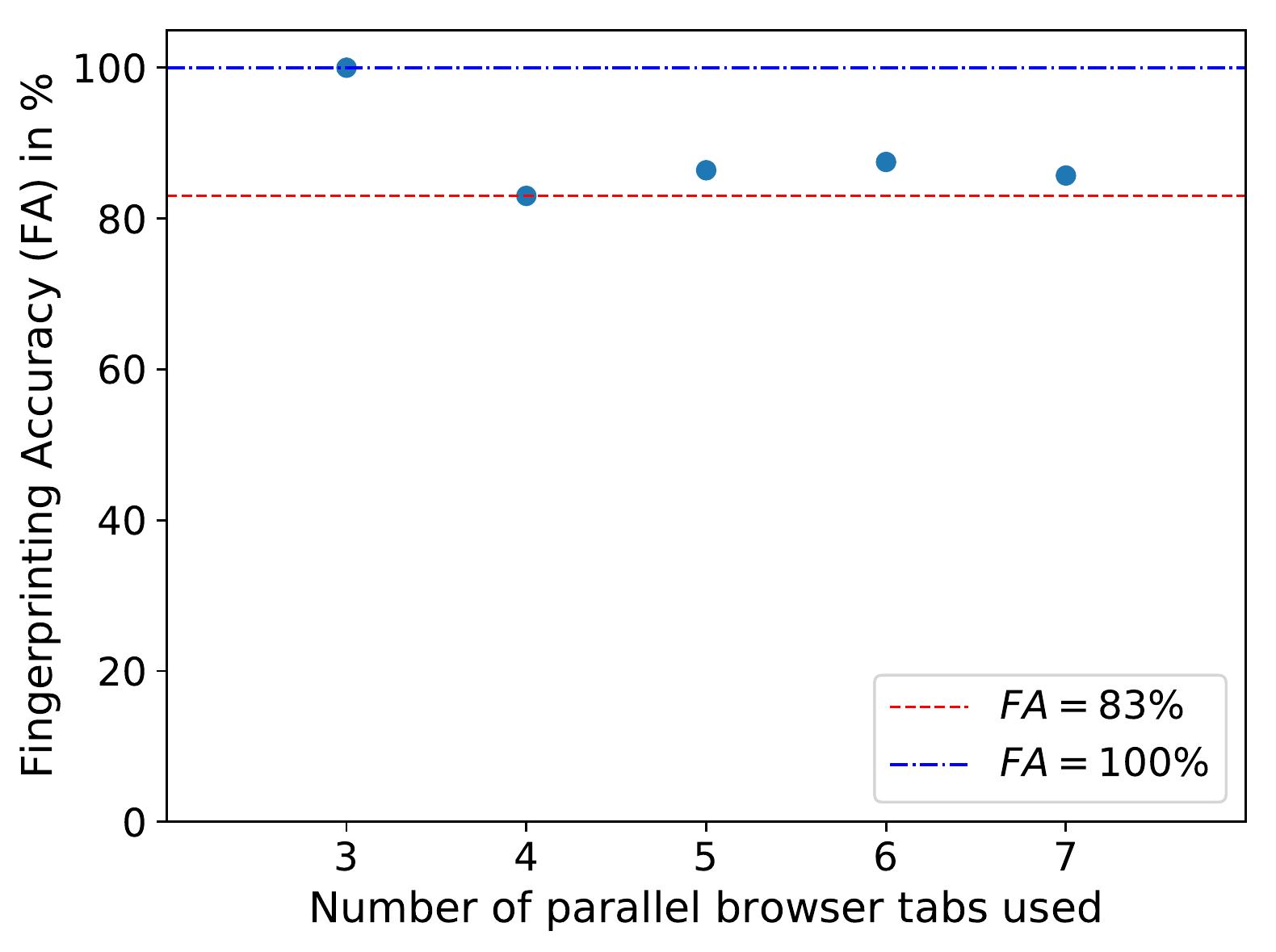}
    \caption{Fingerprinting accuracy ($FA$) of Snoopy in multi-tab browsing scenario (for website $B\_4$)}
    \label{fig:multitab-result}
\end{figure}

Figure~\ref{fig:multitab-result} shows average the fingerprinting accuracy ($FA$) of Snoopy for browsing sessions for the website $B\_4$, involving different number of parallel browser tabs. We observe that even for $7$ parallel browser tabs, the average fingerprinting accuracy of Snoopy is more than $85\%$.

The traffic samples collected for building the training dataset for Snoopy in this case correspond to single-tab single-webpage browsing sessions. This is in contrast to existing works~\cite{chapman2011automated,danezis2010traffic,miller2014know, xu2018multi} that fingerprint sequences of webpage browsing traffic instead of single webpage accesses. While existing techniques have been shown to result in a high fingerprinting accuracy ($FA$), it magnifies the number of website accesses required ($N_q$) to a great extent. Table~\ref{tab:traces} shows a comparison of the number of webpage sequences of length $3-7$ for the $10$ websites we selected for our experiments, with the number of individual webpages that Snoopy would access for building fingerprints. Existing works would require $N_q = l\times$ number of traffic samples for training their models, where $l$ is the number of webpage sequences possible. On the other hand, Snoopy would require only $N_q = w\times$ number of traffic samples, where $w$ $(\ll l)$ is the number of webpages in the website. Note that the number of webpage sequences shown in Table~\ref{tab:traces} do not consider multiple occurrences of the same webpage in a sequence. Considering such cases would have further increased $N_q$ for existing works. This might have caused the website to block the adversary from collecting traffic samples, and would have definitely escalated the time required for collecting traffic samples, and training the model. On the other hand, this would not impact the $N_q$ value for Snoopy. Latest works~\cite{gu2015novel} that are closest to Snoopy in terms of working principle have not been able to obtain a considerable $FA$ value beyond $2$ parallel tabs.

\begin{table}[t]
\centering
\begin{tabular}{|c|c|c|l|}
\hline
\rowcolor[HTML]{EFEFEF}  \multicolumn{1}{|l|}{\cellcolor[HTML]{EFEFEF}\scriptsize{Website}} & \multicolumn{1}{c|}{\cellcolor[HTML]{EFEFEF}\scriptsize{\begin{tabular}[c]{@{}c@{}}Number of unique webpage \\sequences (Sequence Length: 3-7)\\($l$)\end{tabular}}} & \multicolumn{1}{c|}{\cellcolor[HTML]{EFEFEF}\scriptsize{\begin{tabular}[c]{@{}c@{}}Minimum number of website\\ accesses required by Snoopy\\($w$)\end{tabular}}} \\ \hline
IC\_1 & 586 & 95 \\ \hline
IC\_2 & 170,538 & 849 \\ \hline
B\_1 & 9,291,522 & 444 \\ \hline
B\_2 & 1199 & 458  \\ \hline
B\_3 & 13,617 & 549 \\ \hline
B\_4 & 7,000 & 965 \\ \hline
B\_5 & 103,116 & 1964 \\ \hline
SBC & 3,271 & 27 \\ \hline
RS & 1,159 & 20 \\ \hline
PS & 11 & 11 \\ \hline
\end{tabular}
\caption{Snoopy vs. Existing works: Number of website accesses required for fingerprinting for predicting multi-tab browsing}
\label{tab:traces}
% \vspace{-15pt}
\end{table}

\subsubsection{Generalization across various Operating Systems}\label{sec:os}

\begin{table}[t]
\centering
\begin{tabular}{|c|c|c|}
\hline
\rowcolor[HTML]{EFEFEF} 
\cellcolor[HTML]{EFEFEF}{\color[HTML]{000000} } & \multicolumn{2}{c|}{\cellcolor[HTML]{EFEFEF}{\color[HTML]{000000} Operating System}} \\ \cline{2-3} 
\rowcolor[HTML]{EFEFEF} 
\multirow{-2}{*}{\cellcolor[HTML]{EFEFEF}{\color[HTML]{000000} Classifier}} & {\color[HTML]{000000} \begin{tabular}[c]{@{}c@{}}Training: Windows\\ Testing: Ubuntu\end{tabular}} & {\color[HTML]{000000} \begin{tabular}[c]{@{}c@{}}Training: Ubuntu\\ Testing: Ubuntu\end{tabular}} \\ \hline
\textbf{SNOOPY} & \textbf{97.8} & \textbf{97.9} \\ \hline
ML\_Wfin~\cite{yan2018feature} & 54.8 & 93.7 \\ \hline
ML\_CUMUL~\cite{panchenko2016website} & 21.5 & 78 \\ \hline
ML\_KFP~\cite{hayes2016k} & 15.0 & 65.2 \\ \hline
ML\_BoG~\cite{miller2014know} & 2.1 & 77.9 \\ \hline
ML\_PS~\cite{cai2012touching,chapman2011automated} & 1.1 & 77.9 \\ \hline
ML\_IPS~\cite{cai2012touching,chapman2011automated} & 1.1 & 62.1 \\ \hline
ML\_OPS~\cite{cai2012touching,chapman2011automated} & 1.1 & 55.8 \\ \hline
ML\_LL~\cite{liberatore2006inferring} & 1 & 40.0 \\ \hline
\end{tabular}
% \vspace{5pt}
\caption{Accuracy (in \%) of Snoopy vs. ML models when tested on data points from a different operating system (for 95 webpages).} 
\label{tab:os}
\end{table}

\hlt{We now evaluate Snoopy and existing works on their ability to generalize across different operating systems ($O$). In this experiment we consider the website $IC\_1$ and $GF=\{I,BC,T,O,B,N\}$, such that $I$ is the set of $95$ webpages of the website $IC\_1$, $BC$ = $\{\{$Caching OFF, Cookies Allowed$\}\}$, $T=1$, $B=\{Firefox\}$, and $N$ was constant. We built two different training datasets, one where $O=\{Ubuntu\}$ was kept constant, and another where $O=\{Windows\}$ was kept constant. For building each training dataset, we collected $10$ samples of browsing traffic from each of the webpages in $I$. For this experiment, $N_q=95\times 10$, which is the constraint imposed by the finite query model. In our test dataset, we considered $O=\{Ubuntu\}$, with all other factors similar to the training datasets.}

% We now evaluate Snoopy and existing works on their ability to generalize across different operating systems ($O$). In this experiment we consider the website $IC\_1$. For building our test dataset for this experiment, we considered $200$ traffic traces having generalization factors $GF=\{I,BC,T,O,B,N\}$, such that $I$ is the set of $95$ webpages of the website $IC\_1$, $T=1$, $O=\{Ubuntu\}$, $B=\{Firefox\}$, $BC$ = $\{\{$Caching OFF, Cookies Allowed$\}\}$, and $N$ was constant. We built two different training datasets -- one using traffic samples generated using Windows OS and another using samples generated using Ubuntu OS. Each dataset comprised $10$ single-tab browsing traffic traces from each of the $95$ webpages. 

% Table~\ref{tab:os} shows that while Snoopy achieves a fingerprinting accuracy ($FA$) of $97.8\%$, the $FA$ values for existing ML-based techniques were much lower (maximum $54.8\%$) when they were trained with traffic samples from Windows for predicting webpages accessed from a different OS, i.e., Ubuntu.
% However, the $FA$ values for ML-based techniques improved significantly when both the training traffic samples and the test traffic samples belonged to the same OS (for e.g., $FA$ for ML\_Wfin~\cite{yan2018feature} increased from $54.8\%$ to $93.7\%$). 

\hlt{Table~\ref{tab:os} shows a comparison of the fingerprinting accuracy ($FA$) of Snoopy and related works. When training and testing were performed using traffic samples from the same OS, Snoopy achieves an $FA$ that is comparable ($FA\approx 94\%-98\%$) to the best performing ML model ML\_Wfin~\cite{yan2018feature}, as expected. However, when the OS used for testing and training were different, Snoopy outperforms even the best ML techniques by a huge margin. For example, Snoopy achieves an $FA=97.8\%$ as compared to ML\_Wfin~\cite{yan2018feature} that achieves $FA=54.8\%$.}

\subsubsection{Generalization across various Browsers}\label{sec:browser}

\begin{table}[t]
\centering
\begin{tabular}{|c|c|c|}
\hline
\rowcolor[HTML]{EFEFEF} 
\cellcolor[HTML]{EFEFEF}{\color[HTML]{000000} } & \multicolumn{2}{c|}{\cellcolor[HTML]{EFEFEF}{\color[HTML]{000000} Browser}} \\ \cline{2-3} 
\rowcolor[HTML]{EFEFEF} 
\multirow{-2}{*}{\cellcolor[HTML]{EFEFEF}{\color[HTML]{000000} Classifier}} & {\color[HTML]{000000} \begin{tabular}[c]{@{}c@{}}Training: Chrome\\ Testing: Firefox\end{tabular}} & {\color[HTML]{000000} \begin{tabular}[c]{@{}c@{}}Training: Firefox\\ Testing: Firefox\end{tabular}} \\ \hline
\textbf{SNOOPY} & \textbf{85.9} & \textbf{97.9} \\ \hline
ML\_Wfin~\cite{yan2018feature} & 13.8 & 93.7 \\ \hline
ML\_PS~\cite{cai2012touching,chapman2011automated} & 8.0 & 77.9 \\ \hline
ML\_OPS~\cite{cai2012touching,chapman2011automated} & 8.0 & 55.8 \\ \hline
ML\_IPS~\cite{cai2012touching,chapman2011automated} & 5.7 & 62.1 \\ \hline
ML\_LL~\cite{liberatore2006inferring} & 5.7 & 40.0 \\ \hline
ML\_CUMUL~\cite{panchenko2016website} & 3.4 & 78 \\ \hline
ML\_BoG~\cite{miller2014know} & 2.3 & 77.9 \\ \hline
ML\_KFP~\cite{hayes2016k} & 2.3 & 65.2 \\ \hline
\end{tabular}
% \vspace{5pt}
\caption{Accuracy (in \%) of Snoopy vs. ML models when tested on data points from a different browser (for 95 webpages).} 
\label{tab:browser}
\end{table}

\hlt{We now evaluate Snoopy and existing works on their ability to generalize across different browsers ($B$). In this experiment we consider the website $IC\_1$ and $GF=\{I,BC,T,O,B,N\}$, such that $I$ is the set of $95$ webpages of the website $IC\_1$, $BC$ = $\{\{$Caching OFF, Cookies Allowed$\}\}$, $T=1$, $O=\{Ubuntu\}$, and $N$ was constant. We built two different training datasets, one where $B=\{Google\ Chrome\}$ was kept constant, and another where $B=\{Firefox\}$ was kept constant. For building each training dataset, we collected $10$ samples of browsing traffic from each of the webpages in $I$. For this experiment, $N_q=95\times 10$, which is the constraint imposed by the finite query model. In our test dataset, we considered $B=\{Firefox\}$, with all other factors similar to the training datasets.}

% We now evaluate Snoopy and existing works on their ability to generalize across different browsers. In this experiment we consider the website $IC\_1$. For building our test dataset for this experiment, we considered $200$ traffic traces having generalization factors $GF=\{I,BC,T,O,B,N\}$, such that $I$ is the set of $95$ webpages of the website $IC\_1$, $T=1$, $O=\{Ubuntu\}$, $B=\{Firefox\}$, $BC$ = $\{\{$Caching OFF, Cookies Allowed$\}\}$, and $N$ was constant. We built two different training datasets -- one using traffic samples generated using Google Chrome browser and another using samples generated using Firefox browser. Each dataset comprised $10$ single-tab browsing traffic traces from each of the $95$ webpages.

\hlt{Table~\ref{tab:browser} shows a comparison of the fingerprinting accuracy ($FA$) of Snoopy and related works. When training and testing were performed using traffic samples from the same browser, Snoopy achieves an $FA$ that is comparable ($FA\approx 94\%-98\%$) to the best performing ML model ML\_Wfin~\cite{yan2018feature}, as expected. However, when the OS used for testing and training were different, Snoopy outperforms even the best ML techniques by a huge margin. For example, Snoopy achieves an $FA=85.9\%$ as compared to ML\_Wfin~\cite{yan2018feature} that achieves $FA=13.8\%$.}

\subsubsection{Generalization across network conditions}\label{sec:network}
\hlt{We did not observe any changes in the fingerprinting accuracy of Snoopy due to small-scale natural network fluctuations. The results shown in Table~\ref{table:predaccuracy1} reflects the prediction accuracy of Snoopy across diverse network conditions. The traffic samples that were used for profiling were collected over a steady network connection within a short span of time. On the contrary, the test traffic traces were collected over a month from different geographical regions to ensure sufficient diversity in network conditions. 

To further test the limits of Snoopy, we introduced artificial perturbations in the compromised network device over an hour by randomly adding delays of 50ms to 80ms and throttling the bandwidth by $20\%$. In this case, we encountered significant packet drops. This resulted in incomplete download of web resources on several instances, and consequently Snoopy could not identify the webpages correctly. The percentage of cases where the webpages could not be identified by Snoopy has been indicated in Table~\ref{table:predaccuracy1}.}     

\subsubsection{Resource-level prediction accuracy of Snoopy}\label{sec:resource-accuracy}
\begin{table}[t]
\centering
\begin{tabular}{|c|c|c|c|c|}
\hline
\rowcolor[HTML]{EFEFEF} 
\cellcolor[HTML]{EFEFEF} & \cellcolor[HTML]{EFEFEF} & \multicolumn{3}{c|}{\cellcolor[HTML]{EFEFEF}\scriptsize\textbf{Resource identification accuracy}} \\ \cline{3-5} 
\rowcolor[HTML]{EFEFEF} 
\cellcolor[HTML]{EFEFEF} & \cellcolor[HTML]{EFEFEF} & \cellcolor[HTML]{EFEFEF} & \multicolumn{2}{c|}{\cellcolor[HTML]{EFEFEF}\scriptsize\textbf{\begin{tabular}[c]{@{}c@{}}Unidentified resources\end{tabular}}} \\ \cline{4-5} 
\rowcolor[HTML]{EFEFEF} 
\multirow{-3}{*}{\cellcolor[HTML]{EFEFEF}\scriptsize\textbf{Website}} & \multirow{-3}{*}{\cellcolor[HTML]{EFEFEF}\scriptsize\textbf{\begin{tabular}[c]{@{}c@{}}No of \\ resources\end{tabular}}} & \multirow{-2}{*}{\cellcolor[HTML]{EFEFEF}\scriptsize\textbf{\begin{tabular}[c]{@{}c@{}}Accurately \\ identified \\ (\%)\end{tabular}}} & \scriptsize\textbf{\begin{tabular}[c]{@{}c@{}}Incomplete \\ download\\ (\%)\end{tabular}} & \scriptsize\textbf{\begin{tabular}[c]{@{}c@{}}Conflict\\ (\%)\end{tabular}} \\ \hline
IC\_1 & 187 & 94 & 5 & 1 \\ \hline
IC\_2 & 19,163 & 71 & 19 & 10 \\ \hline
B\_1 & 865 & 69 & 7 & 24 \\ \hline
B\_2 & 12,435 & 47 & 52 & 1 \\ \hline
B\_3 & 688 & 91 & 5 & 4 \\ \hline
B\_4 & 3472 & 83 & 1 & 16 \\ \hline
B\_5 & 3,998 & 68 & 26 & 6 \\ \hline
SBC & 277 & 81 & 7 & 12 \\ \hline
RS & 204 & 94 & 6 & 0 \\ \hline
PS & 55 $(11^{\#})$ & 15 $(76^{\#\#})$ & - & 0 \\ \hline
\end{tabular}
\caption{Prediction accuracy of resources in bot traces\\
\scriptsize{$^{\#}$The number indicates the resources that are of interest to the adversary}\\
\scriptsize{$^{\#\#}$Accuracy computed with respect to the number of resources of interest}}
% \caption{Prediction accuracy of resources in bot traces\\ {\scriptsize*(U-F)-Ubuntu-Firefox, **(W-F)-Windows-Firefox, ***(U-C)-Ubuntu-Chrome}}
\label{table:predaccuracy}
% \vspace{-10pt}
\end{table}
We now present more detailed evaluation results about the performance of Snoopy. This includes individual resource-level prediction accuracy of Snoopy and the correlation between resource-level prediction accuracy and webpage-level prediction accuracy. We also analyze the cases where Snoopy fails to accurately identify the webpages accessed, and propose potential solutions to effectively handle such scenarios.

Table~\ref{table:predaccuracy} shows the resource prediction accuracy of Snoopy on the ten websites. While for most of the websites Snoopy had a prediction accuracy of more than $70\%$, in a few cases, we witnessed an accuracy of less than $50\%$. For websites with a very low number of resources (for instance, IC\_1 and RS) Snoopy had a high prediction accuracy since the encrypted size of most of the resources were very distinct from each other. On the other hand, Snoopy had a low prediction accuracy for most websites with a high number of resources (for instance, B\_2 and B\_5). This is because, in practice, most of the web resources in such websites have similar sizes. However, the website IC\_2 was an exception where Snoopy had a relatively high prediction accuracy despite a large number of resources. This was because, most of the resources in this website had distinct sizes.  
Also, note that out of the $55$ resources in the website PS, only $11$ resources were of interest to the adversary, and their download sequence was sufficient to identify all the $40323$ webpages uniquely. Details about this can be found in our recent work~\cite{mitra2020depending}.

\subsubsection{Relation between resource prediction accuracy and webpage prediction accuracy}\label{sec:resvspage}
As seen from Table~\ref{table:predaccuracy1} and Table~\ref{table:predaccuracy}, the relationship between web resource prediction and webpage prediction accuracy is not straightforward. We now discuss three different scenarios that we encountered during the evaluation of Snoopy.

\noindent\textbf{Case 1: High resource prediction accuracy and high webpage prediction accuracy.} In case of websites IC\_1 and B\_3, we observed a high correlation between web resource prediction accuracy ($91\%-94\%$) and webpage prediction accuracy ($93\%-97\%$). While such results are quite intuitive, such direct correlation was not observed in case of some other websites;

\noindent\textbf{Case 2: Low resource prediction accuracy and high webpage prediction accuracy.}  \hltr{In case of IC\_2, B\_1, B\_2 and B\_5, we see low values of resource prediction accuracy ($47\%-75\%$) but relatively high values of webpage prediction accuracy ($89\%-99\%$). The website B\_2 exhibits an extreme case of this behavior with only $47\%$ resource prediction accuracy but $99\%$ webpage prediction accuracy. 
The reason for this is though the overall resource prediction accuracy was low, Snoopy was able to identify the critical resources that are unique to a webpage; and,}

\noindent\textbf{Case 3: High resource prediction accuracy and low webpage prediction accuracy.} In case of the website RS, though the resource prediction accuracy was high ($94\%$), the webpage prediction accuracy was relatively low ($75\%$). This was because most of the web resources that Snoopy could predict correctly were non-critical resources that were shared by multiple webpages. 

\noindent\textbf{Failure Analysis} -- Our analysis reveals that \hltr{the most common reasons due to which Snoopy could not perform well} are \textbf{(1)} when the identified resources are associated with a lot of other webpages;  \textbf{(2)} when the resources critical for identifying a webpage are already cached at the browser and do not get downloaded; and, \textbf{(3)} increase in number of possible webpage transitions in case of multi-tab browsing. For instance, Snoopy may correctly identify a web resource but may not be able to determine which tab it came from in case webpages open on multiple tabs at the same time share the resource. This increases the confusion and leads to incorrect predictions. \hltr{To improve the webpage prediction accuracy in cases where Snoopy has a poor accuracy, we explore an ensemble of Snoopy and a ML-based technique in Section~\ref{sec:snoopy-meets-ml}, that complies with the constraints of a finite query model.}

%%%%%%%%%%%%%%%%%%%%%%%%%%%%%%%%%%%%%%%%%%%%%%%%%%%%%%%%%%%%%%
%%%%%%%%%%%%%%%%%%%%%%%%%%%%%%%%%%%%%%%%%%%%%%%%%%%%%%%%%%%%%%
\subsection{Snoopy-ML Ensemble}\label{sec:snoopy-meets-ml}
Our experiments (Refer to Section~\ref{results}) show that there are certain cases where the webpage prediction accuracy of Snoopy is approximately $80\%$. This is comparatively lower than most of the other websites where Snoopy could achieve a prediction accuracy of more than $90\%$. This motivated us to design a small experiment where we analyze the effectiveness of ML-based webpage identification techniques in classifying only those webpages that Snoopy fails to classify.
\subsubsection{Experiment 1}\label{sec:ensemble-1}
For our experiment we consider the website B\_4, and the scenario described in Section~\ref{sec:user-interests}, where we study the generalization capability of Snoopy and ML-based techniques in terms of user interests. We had observed that for $200$ webpages of website B\_4 and $10$ traffic samples per webpage, Snoopy had achieved a prediction accuracy of $78\%$. We first examine how the best-performing ML-based technique, ML\_Wfin~\cite{yan2018feature}, performs when trained and tested on only those webpages that Snoopy failed to identify.  

We first inspect the prediction results of Snoopy and identify $70$ out of $200$ webpages that Snoopy could not predict correctly. Next, we train the ML\_Wfin model using $10$ samples from each of these $70$ webpages. During validation, the ML model assigns a probability to each of these $70$ classes (webpages) for each validation point. For a given validation point (say, $t_j$), the webpage that gets assigned the highest probability (denoted as $P_j$) is the prediction output. 
For the $70$ webpages, we observe a validation accuracy of $\approx94\%$. 
% with a minimum prediction probability ($min(P_j), \forall_j$) of $32.5\%$. We denote this probability value as $P_v$, and use it as a threshold value for classification. 
The results of this experiment motivated us to further explore if there is a way to combine Snoopy and an ML-based technique into an ensemble that outperforms each of the aforementioned webpage identification techniques individually; subject to the limitations on the number of queries.

\subsubsection{Experiment 2}
We build a very basic ensemble model with Snoopy and ML\_Wfin~\cite{yan2018feature} as sub-modules. We train (webpage profiling) Snoopy on $130$ webpages of the website B\_4, while we train ML\_Wfin on the remaining $70$ webpages of B\_4. We use $10$ traffic samples from each webpage for this training. 
% Since we are considering the case where generalization is performed only with respect to user interests, we do not need more training samples for ML\_Wfin. However, note that we would have to collect more training samples for ML\_Wfin if we considered other factors in $\mathbb{GF}$ as well.
\begin{figure}[t]
    \centering
    \includegraphics[scale=0.15]{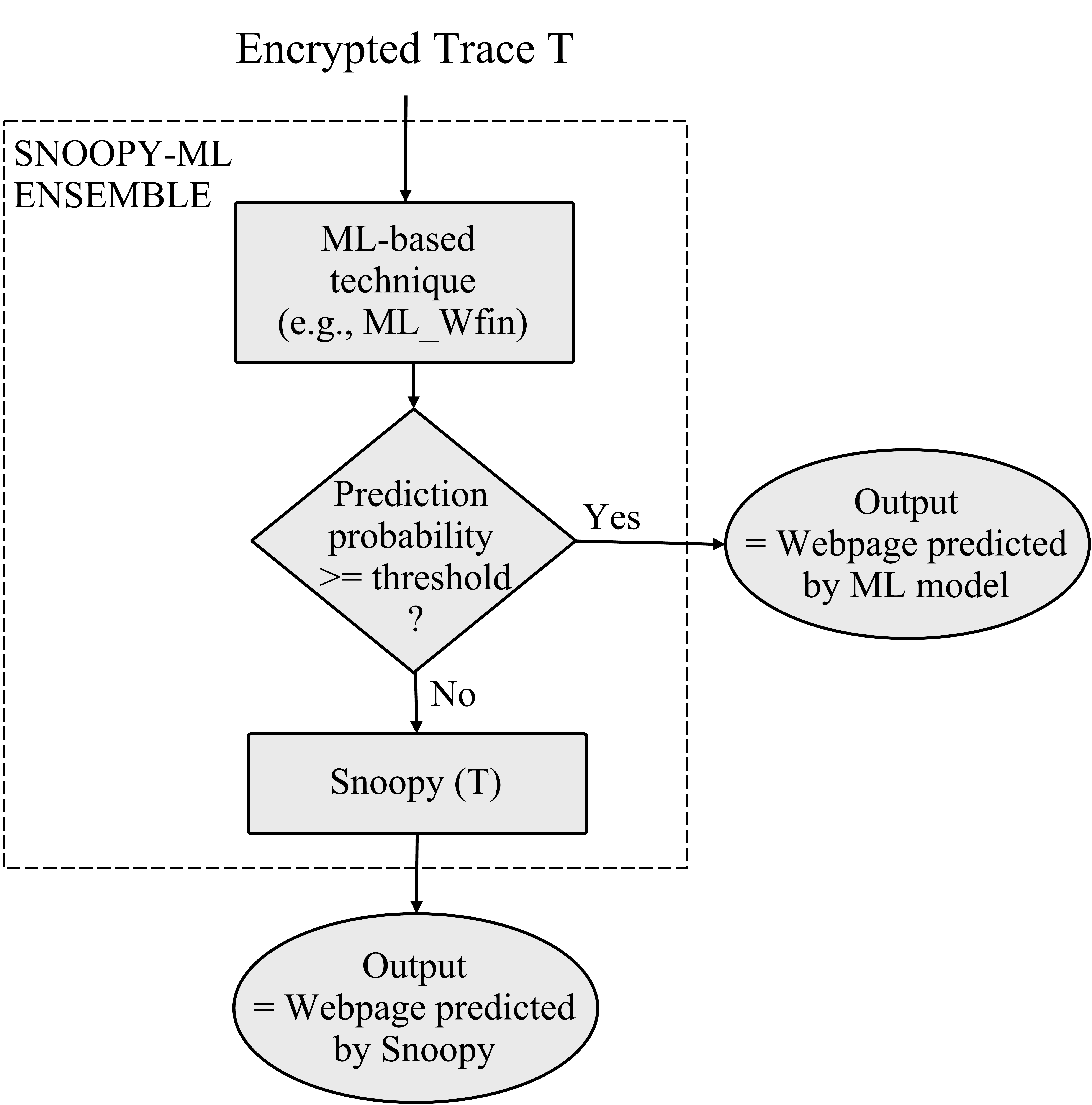}
    % \caption{A High Level Overview of Snoopy\vspace{-15pt}}
    \caption{Webpage prediction by Snoopy-ML ensemble}
    \label{fig:ensemble}
    \vspace{-15pt}
\end{figure}
Figure~\ref{fig:ensemble} shows how this ensemble is used for predicting webpages accessed in a given encrypted trace $T$.
The trace $T$ is first passed to ML\_Wfin for prediction. ML\_Wfin predicts one out of the $70$ webpages it has been trained with, with a probability $P_{ML}$. 
Based on our insights from Experiment 1 (refer to Section~\ref{sec:ensemble-1}), we calculate a threshold probability $P_v=min(P_j), \forall_j$ (where, $P_j$ is the maximum probability assigned to a class by ML\_Wfin for validation point $t_j$) to determine if $T$ belongs to one of the $70$ webpages that ML\_Wfin was trained with. 
% We consider the minimum validation prediction probability of ML\_Wfin, $P_v$ (derived from Experiment 1 in Section~\ref{sec:ensemble-1}), as a threshold to determine if $T$ belongs to one of the $70$ webpages that ML\_Wfin was trained with. 
If $P_{ML} \geq (P_v-10)$ (error margin = $10\%$), we consider the webpage predicted by ML\_Wfin as the ensemble output. If $P_{ML} < (P_v-10)$, the trace $T$ is passed to Snoopy for prediction, and we consider the class predicted by Snoopy as the ensemble output. For the scenario we considered in Experiment 1, the observed value of $P_v$ was $32.5\%$.

For testing this basic ensemble, we use a dataset comprising encrypted traffic traces from the website $B\_4$, collected using the same browsing context as the training dataset. Our ensemble achieved a prediction accuracy of $97\%$, which is much higher than the accuracy achieved by Snoopy ($78\%$) or ML\_Wfin ($76\%$) separately. Since in this experiment we considered the case where generalization is performed only with respect to user interests, we did not need additional training samples for ML\_Wfin. However, note that we would have to collect more training samples for ML\_Wfin if we considered other factors in $\mathbb{GF}$ as well. While this might not be the best possible ensemble, the fact that even such a basic ensemble achieved a high accuracy shows an interesting future research direction.

\section{Conclusion}\label{concl}
% The conclusion goes here.
In this paper, we proposed Snoopy, a webpage fingerprinting framework for performing mass surveillance while complying with a finite query model. 
Snoopy achieves this objective with $\approx90\%$ accuracy when evaluated on most websites, across various browsing contexts, while requiring only $3-10$ traffic samples per webpage. 
We also presented some preliminary findings on the possibility of a Snoopy-ML ensemble model for websites where Snoopy alone could not achieve the best results. In such a case, when we tried to generalize across various user interests, a simple ensemble of Snoopy and an ML-based technique achieved $\approx97\%$ accuracy while adhering to the finite query model. However, the number of traffic samples required by the ML component of the ensemble would still multiply if we attempted to generalize it across all the factors in the browsing context. Therefore, we believe that it is a challenging endeavor to design Snoopy-ML ensembles that work with a limited number of queries, and this could be a future research direction.

\bibliographystyle{IEEEtran}
% \bibliography{references.bib}
%

%%**********************************************************************************************************************************************************
% if have a single :
\clearpage
\appendix[Assessing stability of known traffic features] \label{appendix}

\begin{figure}[t]
    \begin{subfigure}[b]{0.5\textwidth}
    \centering
      \includegraphics[scale=0.35]{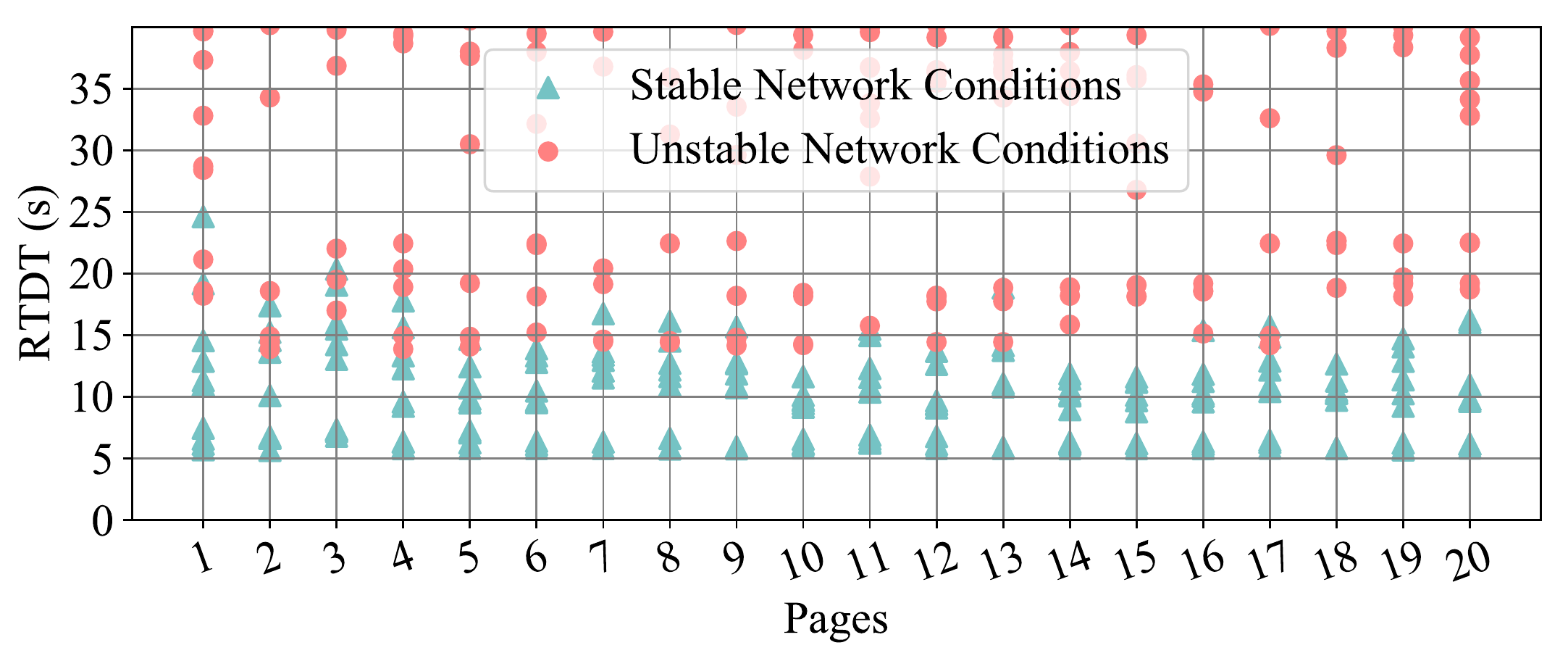}
      \captionsetup{justification=centering}
      \caption{Variation in Round Trip Download Time (RTDT) of webpages in Website RS.}
      \label{fig:rtt}
    \end{subfigure}
    \begin{subfigure}[b]{0.5\textwidth}
    \centering
      \includegraphics[scale=0.52]{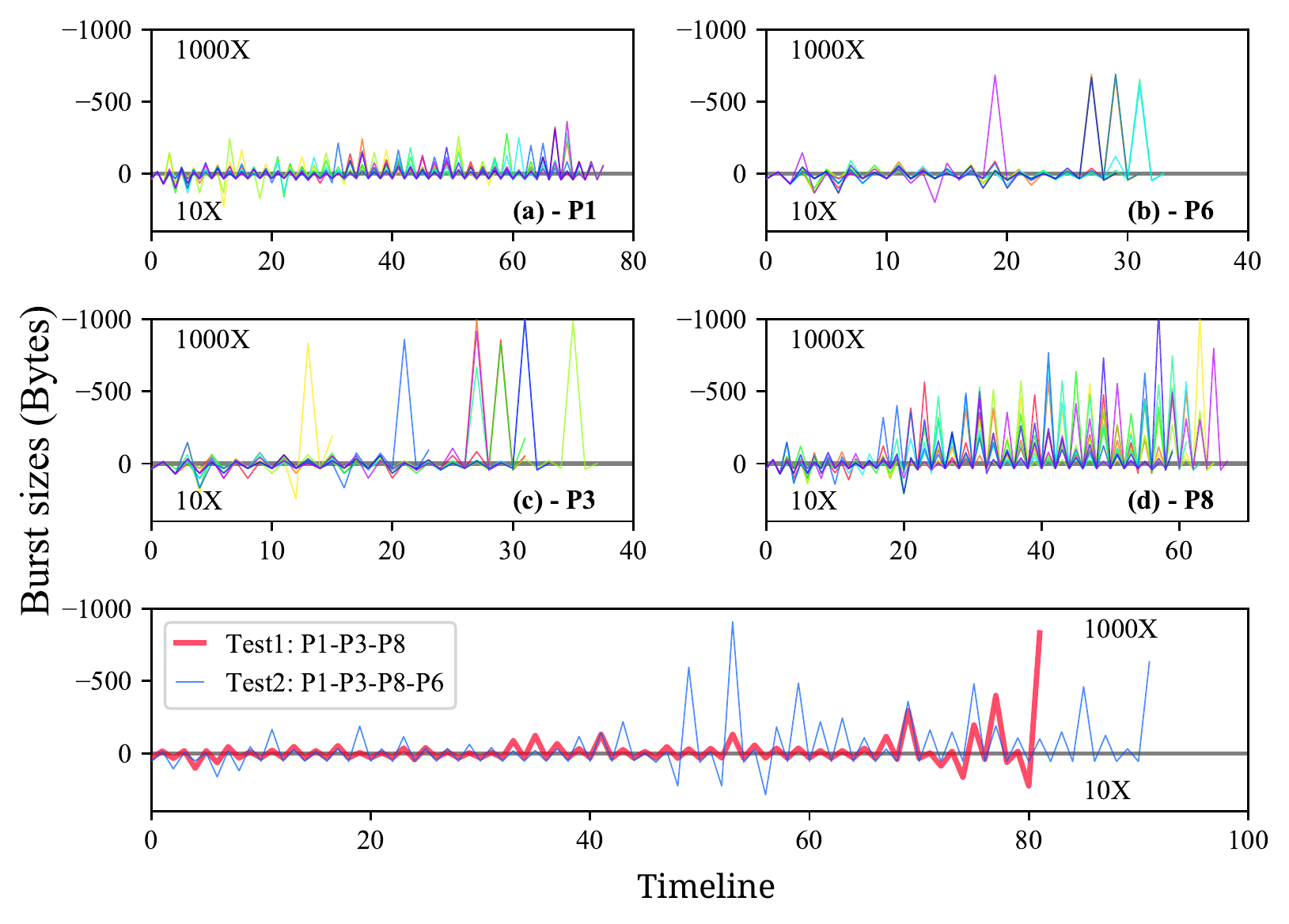}
      \captionsetup{justification=centering}
      \caption{Traffic burst patterns for four webpages in Website RS with caching disabled, and for two different instances of sequential webpage accesses with caching.}
      \label{fig:burst}
    \end{subfigure}
    \caption{Suitability assessment of known side-channels}
    \label{fig:features}
    % \vspace{-15pt}
  \end{figure}

To find a stable side-channel, we assessed the suitability of some widely used simple traffic features such as Round Trip Download Time (RTDT) and packet burst patterns that have been proposed in prior works~\cite{gong2012website,miller2014know}.  
Our analysis presented in Figure~\ref{fig:rtt} shows that the \emph{Round Trip Download Time (RTDT)} of a given webpage varies significantly even with minor fluctuations in network speed (simulated by adding a random per-packet delay of $\approx25$ ms) and hence, not stable. Likewise, packet burst patterns are also not stable. Figure~\ref{fig:burst} shows the packet burst pattern of webpages, namely, P1, P3, P6, and P8, of a website $RS$ (anonymized) over a period of time. The incoming burst sizes are considered to be negative and outgoing ones are considered to be positive. We can observe that the packet burst patterns were different each time we accessed the same webpage. Further, as expected, we could not find any discernible patterns when the webpages were accessed in sequence (e.g., P1-P3-P8), due to the interleaving of client requests and server responses. Such interleaving is quite common in multi-tab browsing, thereby making this side-channel not suitable. Interestingly, these side-channels worked well in the context of targeted surveillance. This is perhaps because the network conditions were predictable and the browsing habits of the target were known.

%%%%%%% This part was previously commented out %%%%%%%%%%%%%
% packet burst sizes for four different webpages of an anonymized website $RS$ across multiple accesses under stable network conditions.  The packet burst sizes vary across multiple accesses even for the same webpage.

% Figure~\ref{fig:burst} also shows the packet burst sizes when the webpages were accessed sequentially within the same browsing session. It can be observed that, due to the interleaving of client requests and server responses, which is more prominent in the case of multi-tab browsing, it becomes challenging to extract packet bursts for individual web resources. Thus, packet burst size is also not a stable side-channel.

% Another assumption made by existing works is that the network and browsing conditions when \textit{building} the model are similar to those when \textit{using} the model. For instance, for RTDT to work, the network conditions during profiling must be the same as that of the real users of the website. While this assumption might be useful in case of targeted surveillance, it results in these solutions being inapplicable for mass surveillance, wherein the prediction model is built once and used for various users at various points in time.
%%%%%%% End of commented out part %%%%%%%%%%%%%

\begin{figure}[t]
\begin{subfigure}{.5\textwidth}
\centering
  \includegraphics[scale=0.05]{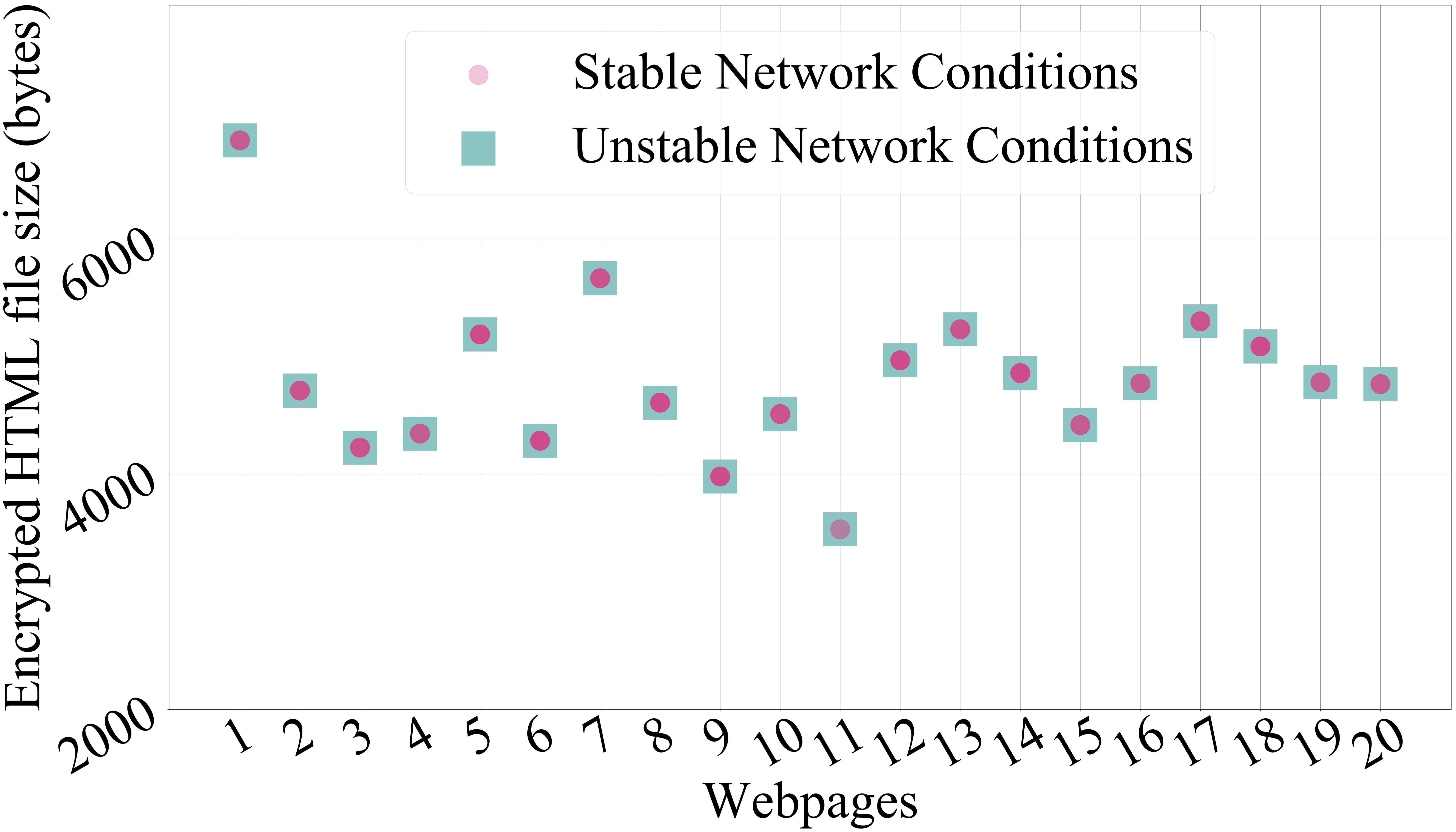}
    \captionsetup{justification=centering}
    \caption{Encrypted resource sizes under stable and varying network conditions}
    \label{fig:pagesizes}
\end{subfigure}
% \hfill
\begin{subfigure}{.5\textwidth}
  \centering
  \includegraphics[scale=.35]{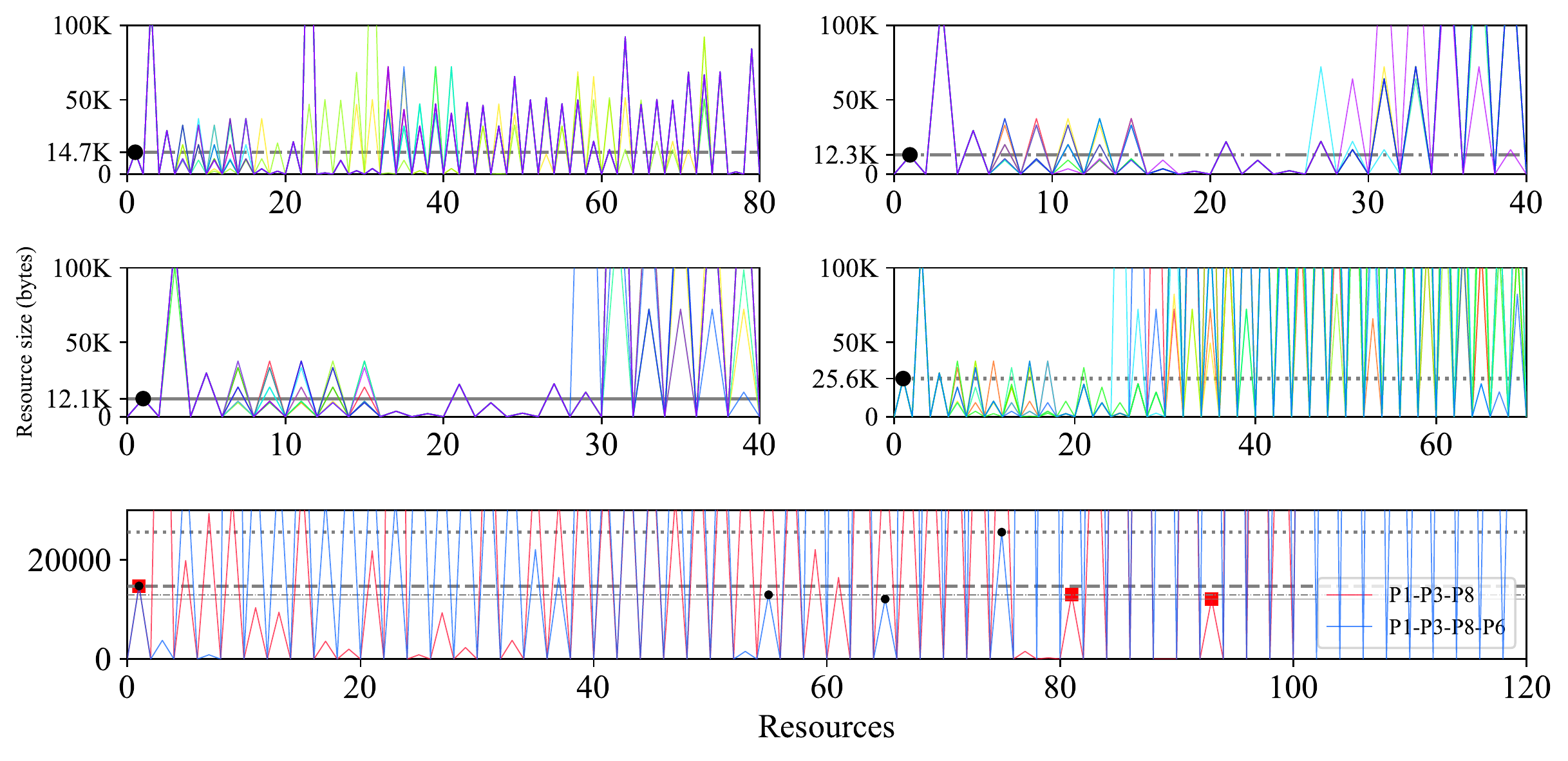}
%     \captionsetup{justification=centering}
    \caption{Encrypted resource sizes for four individual and two sequential webpage accesses. The black dots and red squares denote encrypted HTML file sizes in the two traces. Test 1 considers single-tab browsing and Test 2 considers multi-tab browsing.}
    \label{fig:htmlr} 
\end{subfigure}
\caption{Stability of encrypted resource sizes for a HTTP/1.1 website}
\label{fig:http1stability}
% \vspace{-15pt}
\end{figure}

We now analyse the stability of encrypted resource sizes of webpages in a website as a side-channel. Figure~\ref{fig:http1stability} shows the encrypted sizes of a subset of resources of the website $RS$ (using HTTP/1.1) under different network conditions with multi-tab browsing enabled. We can observe that the encrypted resource sizes are fairly consistent. Note that, if a side-channel is stable under multi-tab browsing conditions, then it will also be stable under single-tab browsing conditions. 

\begin{figure}[!t]
\begin{subfigure}{0.5\textwidth}
\centering
  \includegraphics[scale=0.05]{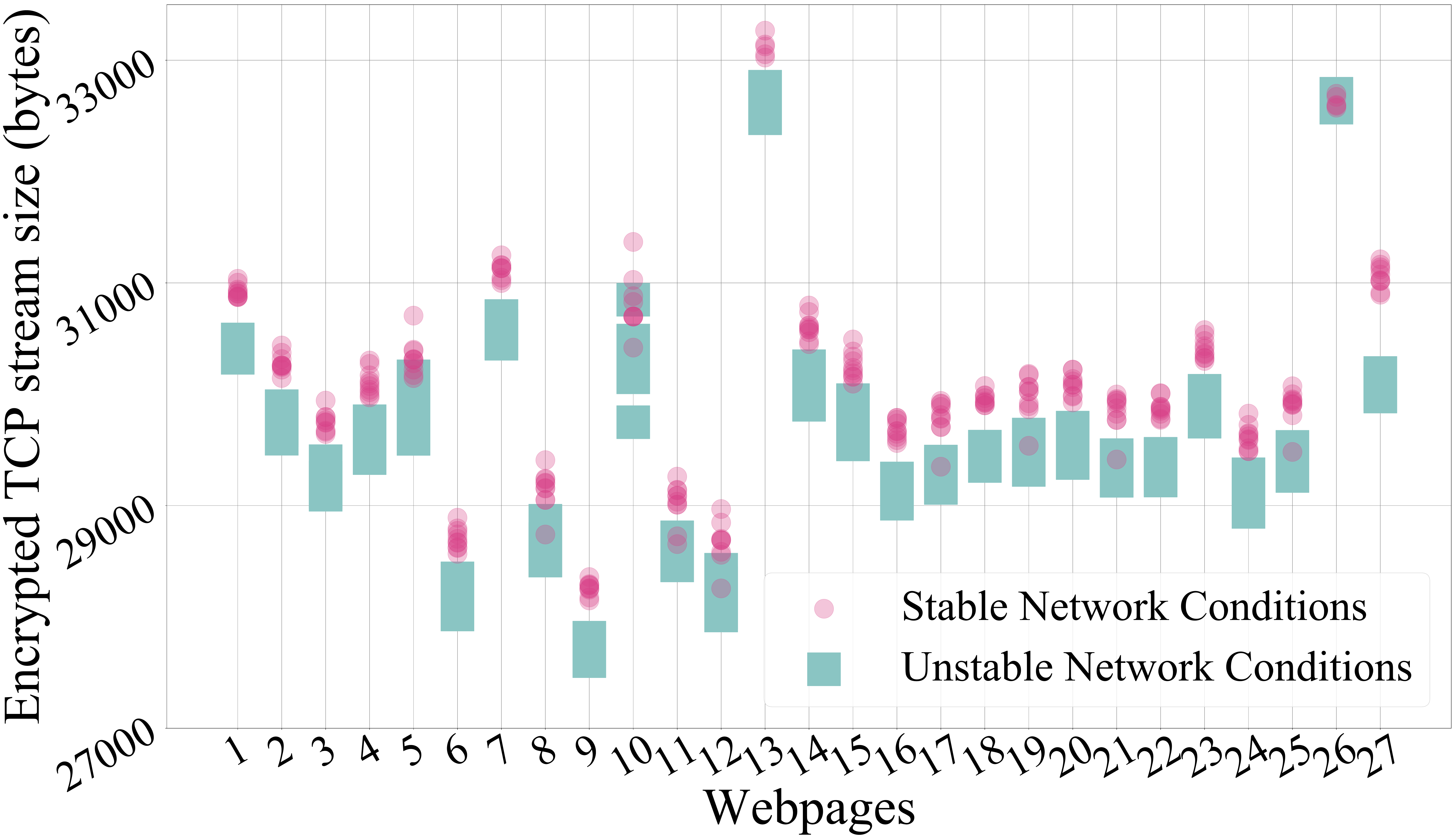}
  \captionsetup{justification=centering}
    \caption{Encrypted TCP stream sizes carrying HTML files under stable and varying network conditions}
    \label{fig:http2ntk}
\end{subfigure}
\begin{subfigure}{0.5\textwidth}
\centering
  \includegraphics[width=6cm,keepaspectratio]{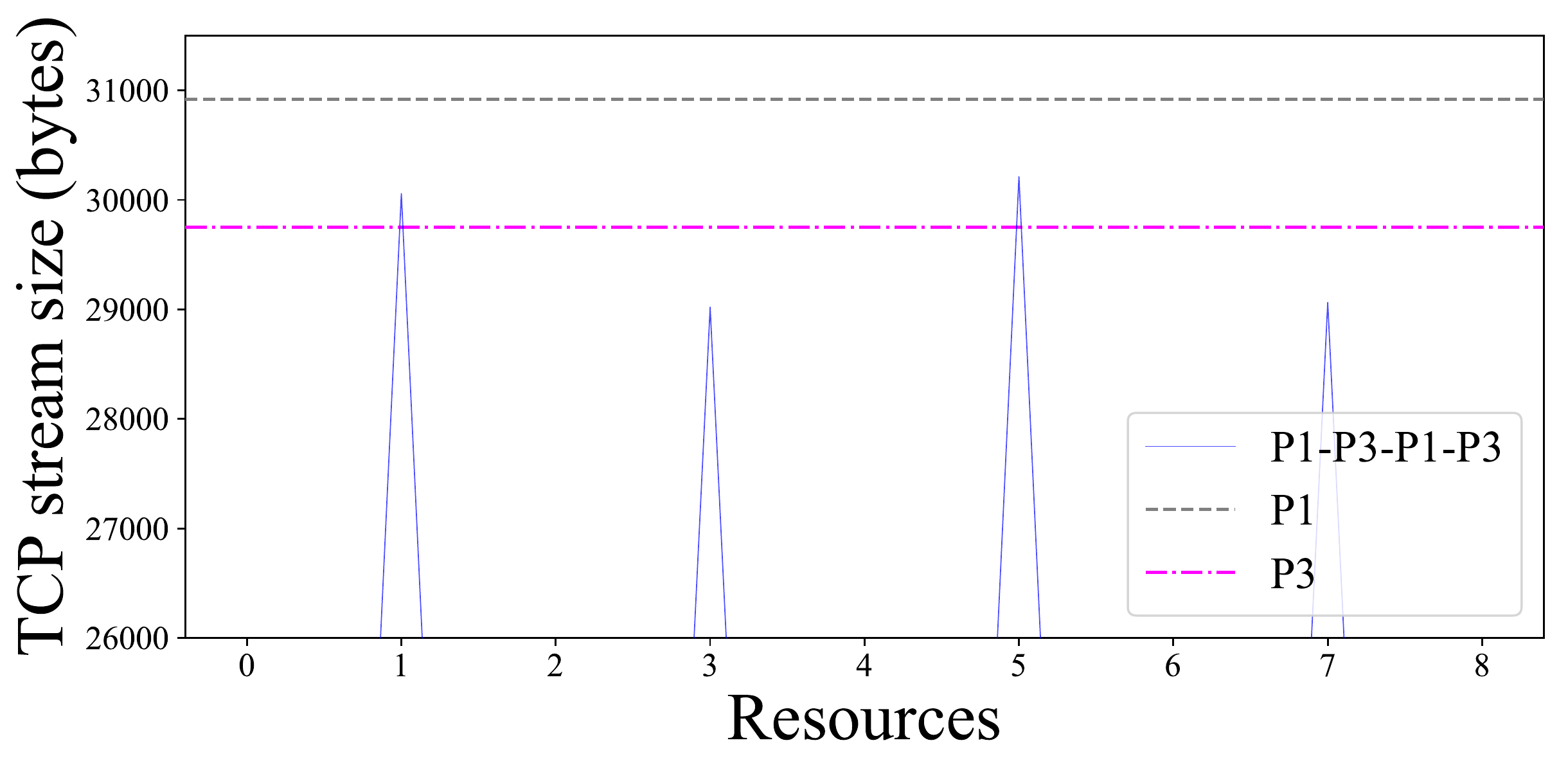}
  \captionsetup{justification=centering}
    \caption{Encrypted TCP stream sizes for two individual and one sequential webpage access using multi-tab browsing.}
    \label{fig:http2multi} 
\end{subfigure}
\caption{Stability of encrypted TCP stream sizes for a HTTP/2 website}
\label{fig:http2stability}
% \vspace{-20pt}
\end{figure}

However, HTTP/2 does not have the notion of encrypted resource sizes. In HTTP/2, many resources could be multiplexed into a single TCP stream, thereby obfuscating the signatures at an individual resource level. In such a case, we consider the size of the TCP stream carrying the resource as a side-channel. Figure~\ref{fig:http2stability} shows the stability of encrypted TCP stream sizes of an anonymized website $SBC$ that uses HTTP/2. Further, we can observe that the TCP stream sizes of the webpages when accessed sequentially are slightly lesser than those corresponding to individual accesses. This happens due to variations in resource sizes caused by tracking and session cookies. An alternative way to fingerprint HTTP/2 webpages is to introduce systematic network delays, throttle the bandwidth or drop packets at the compromised network device, as discussed in one of our prior works~\cite{mitra2020depending}. This would force the web server to transmit the web resources in non-multiplexed fashion, and the attacker can use the same webpage profiling and prediction techniques as those used for HTTP/1.x websites. In this paper, we have used the techniques shown in our prior work~\cite{mitra2020depending} for HTTP/2 websites, to evaluate the performance of the same traffic feature for both HTTP/1.x and HTTP/2 websites.
  
\end{document}